\documentclass[12pt]{iopart}
\usepackage{graphicx}

\begin{document}

\title[Cosmic ray historical building stability monitoring]{Historical building stability monitoring by means of a cosmic ray tracking system}

\author{A Zenoni$^{1,2}$, G Bonomi$^{1,2}$, A Donzella$^{1,2}$,
M Subieta$^{1,2}$, G Baronio$^1$, I Bodini$^1$, D Cambiaghi$^1$, M Lancini$^1$, D Vetturi$^1$, O Barnab\`a$^{3,2}$, F Fallavollita$^{3,2}$,  R Nard\`o$^{3,2}$, C Riccardi$^{3,2}$, M Rossella$^{3,2}$, P Vitulo$^{3,2}$ and G Zumerle$^4$}

\address{$^1$ Universit\`a di Brescia, Via Branze 38, I-25123 Brescia, Italy}
\address{$^2$ Istituto Nazionale di Fisica Nucleare, Via Bassi 6, I-27100 Pavia, Italy}
\address{$^3$ Universit\`a di Pavia, Via Bassi 6, I-27100 Pavia, Italy}
\address{$^4$ Universit\`a di Padova, Via Marzolo 8, I-35131 Padova, Italy}

\ead{aldo.zenoni@unibs.it}

\begin{abstract}
Cosmic ray radiation is mostly composed, at sea level, by high energy muons, which are highly penetrating particles capable of crossing kilometers of rock. Cosmic ray radiation constituted the first source of projectiles used to investigate the intimate structure of matter and is currently and largely used for particle detector test and calibration. 
The ubiquitous and steady presence at the Earth's surface and the high penetration capability has motivated the use of cosmic ray radiation also in fields beyond particle physics, from geological and archaeological studies to industrial applications and civil security.
In the present paper, cosmic ray muon detection techniques are assessed for stability monitoring applications in the field of civil engineering, in particular for static monitoring of historical buildings, where conservation constraints are more severe and the time evolution of the deformation phenomena under study may be of the order of months or years. 
As a significant case study, the monitoring of the wooden vaulted roof of the ``Palazzo della Loggia" in the town of Brescia, in Italy, has been considered. The feasibility as well as the performances and limitations of a monitoring system based on cosmic ray tracking, in the considered case, have been studied by Monte Carlo simulation and discussed in comparison with more traditional monitoring systems.
Requirements for muon detectors suitable for this particular application, as well as the results of some preliminary tests on a muon detector prototype based on scintillating fibers and silicon photomultipliers SiPM are presented.
\end{abstract}

\vspace{2pc}

%
\section{Introduction}\label{Intro}
When primary cosmic rays, mainly composed of high energy protons coming from the sun and from the outer Galaxy, strike the Earth's atmosphere, a cascade of many types of subatomic particles is created~\cite{Beringer2012}. Hadronic particles produced in the shower either interact or decay, and electrons and photons lose energy rapidly through pair production and Bremsstrahlung, so that, by the time the charged component of this particle shower reaches the Earth surface, it comprises primarily positive and negative muons.
The flux reaching the surface of the Earth is about 10,000~$\mu$/(min m$^2$) and the mean muon energy is 3-4~GeV. Since muons are heavy particles and do not undergo nuclear interactions, they are highly penetrating in matter and their average energy is sufficient to penetrate tens of meters of rock.

Cosmic radiation has been known since the first decades of the 20$^{th}$ century and, until the construction of the first particle accelerators, it constituted the best source of projectiles to investigate the fundamental structure of matter and the fundamental interactions between elementary particles.
Nowadays, cosmic rays are largely exploited in nuclear and elementary particle physics for detector testing and calibration and for the alignment of detectors in the very complex apparatuses used in this field~\cite{ALICE2010}.

Making practical use of this natural flux of highly penetrating particles, continuous, free and ubiquitously present on the entire Earth surface, has always been an attractive idea. As the spectrum of cosmic ray muons is continuous and the average range is long, differential attenuation can be used to produce radiographies of large and dense objects. E.P.~George~\cite{George1955} measured, in 1955, the depth of rock above an underground tunnel by making use of the attenuation of cosmic ray muons. 
With the same technique, L.W.~Alvarez performed the radiography of the Second Pyramid of Giza~\cite{Alvarez1970} seeking for the possible presence of hidden chambers. 

Over the following years, muon radiography has been used to perform inspection of large inaccessible systems or even of geographic structures. Several groups are actively working in the imaging of the interior of volcanoes and in the prediction of volcanic eruptions~\cite{Nagamine1995,Tanaka2003,Tanaka2005,Tanaka2007a,Tanaka2007b,Shinohara2012}, \cite{Ambrosi2011,Anastasio2013a,Anastasio2013b}, \cite{Gibert2010,Marteau2012}.
Proposals have been presented to obtain radiographic images of the interior of large vessels with dimensions over many tens of meters, where storage or long-term structural integrity is an important issue~\cite{Jenneson2004,Jenneson2007}. Potential  uses of cosmic ray muon radiography in industrial applications have been explored~\cite{Gilboy2007a,Gilboy2007b, Tanaka2008, Grabski2008}, including the inspection of nuclear waste containers~\cite{Stanley2008} and of the inner structure of a blast furnace~\cite{Nagamine2005,Shinotake2009,Sauerwald2012}. 

In 2003 a new method has been proposed~\cite{Borozdin2003,Priedhorsky2003,Schultz2004,Schultz2007}, the muon tomography, in which the angular scattering that every muon undergoes when crossing matter is exploited. The scattering angles have a Gaussian distribution, with variance proportional to the traversed thickness and to the average ``scattering density" of the material crossed by the muons. The scattering density is roughly proportional to the product of the material mass density times its atomic number. 

This technique needs a more complex apparatus. While the absorption technique requires the measurement of the muon position
and direction only downstream of the object to be inspected, the technique based on muon scattering requires the measurement of muon position and direction both upstream and downstream, to measure the single muon angular deviation. 

This technique has been proposed for the detection of radioactive ``orphan" sources hidden in scrap metal containers~\cite{Pesente2009,Musteel2010,Furlan2013,Benettoni2013}, to inspect commercial cargoes in ports seeking for hidden ``special nuclear materials"~\cite{Riggi2010,Riggi2013a,Riggi2013b,Armitage2013}, to inspect legacy nuclear waste containers~\cite{Mahon2013,Clarkson2013} and to obtain tomographic images of the interior of blast furnaces~\cite{Mublast2013}. In a recent study, the method has been proposed to perform a diagnosis of the damaged cores of the Fukushima reactors~\cite{Borozdin2012}.

In 2007 cosmic ray muon detection techniques were assessed~\cite{Bodini2007} for measurement application in civil and industrial engineering for the monitoring of alignment and stability of large civil and mechanical structures. Situations where environmental conditions are weakly controlled and/or where the pieces whose relative positions are to be monitored are hardly accessible were specifically addressed. A Monte Carlo analysis was developed concerning the case of the alignment of an industrial press. Expected measurement uncertainty and its dependence on the geometry of the set-up, on the presence of materials interposed between the muon detectors and on elapsed time available for the measurement were obtained.

In the present paper, the same general idea of exploiting the cosmic ray natural source of radiation for monitoring of alignment and stability of large structures (muon alignment and stability monitoring) is applied, with an improved detector scheme, to the case of civil buildings. The study is devoted especially to historical buildings, whose cultural and artistic value puts often severe constraints of non-invasiveness to the monitoring techniques that may be employed. 

In particular, the ability of cosmic ray radiation to penetrate large thicknesses of material suffering only small trajectory deviations offers the possibility to overcome the problem of monitoring the relative positions of different points that are physically and optically separated by solid structures as walls or floors. Monitoring systems widely employed, as laser scanner and theodolites, make use of visible light and need complete optical transparency between different reference points. 
In addition, considering that the entire volume of the building is continuously crossed by the flux of the cosmic ray radiation, with directions spanning several tens of degrees from the zenith, a more complex system of muon detectors distributed in different positions of the building may allow a global and simultaneous stability monitoring of the building to be performed.
 
Limiting features of the cosmic ray radiation are the stochastic nature of the deviations of the muon trajectories, due to angular scattering suffered in crossing materials, and the rather low rate of events when detectors of reasonable size are employed. The former feature imposes the use of statistical distributions to extract the measured quantities by statistical inference and, consequently, needs the collection of sufficiently large number of events to obtain adequate measurement precisions. The latter implies the need of rather long data taking times. 

In the case of the monitoring of historical buildings, these negative features might not constitute a severe limitation for the use of the proposed method. Indeed, historical building structures are, in general, characterized by slow processes of deformation, which need to be monitored with fair precision over long periods of time, of the order of months or even years.

\begin{figure}
\begin{center}
\includegraphics[width=0.7\textwidth]{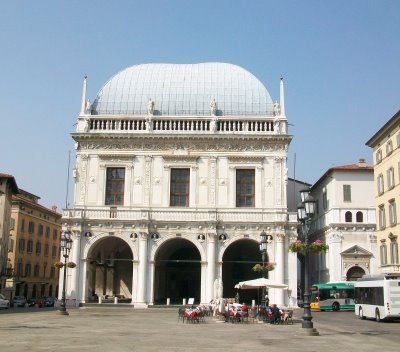}
\caption{\label{Fig1}The ``Palazzo della Loggia" of the town of Brescia (1574).}
\end{center}
\end{figure}

The study of the application of the muon stability monitoring method to historical buildings was performed with a Monte Carlo technique and was applied to a realistic situation, the exemplary case of the ``Palazzo della Loggia", seat of the Mayor, in the town of Brescia, Italy~\cite{Donzella2014}. The ``Palazzo della Loggia" (see~\fref{Fig1}) was built in 1574 by the Venetian Government of the town and suffered, since the first years after the construction, of several structural problems. In recent years, from 1990 to 2001, a campaign of measurements was performed to monitor the stability and progressive deformation of its wooden vaulted roof, completely reconstructed in 1914, by means of a mechanical monitoring system based on the elongation of metallic wires~\cite{Giuriani1993,Giuriani2000}.

In~\Sref{Loggia}, the methodology adopted in the diagnostic phase to understand the static anomalies of the wooden vaulted roof of the ``Palazzo della Loggia" and the results of the analysis are shortly illustrated.
In~\Sref{Muon}, the application of the method of muon stability monitoring to the case of the wooden vaulted roof of the ``Palazzo della Loggia" is described. The measurement system is composed of a number of muon position detectors of given size and given precision that are located in the points of the building structure whose relative positions must be monitored. 

By means of the GEANT4 toolkit for the simulation of the passage of particles through matter~\cite{Agostinelli2003}, the structure of the building (as far as needed) the cosmic ray muon flux, the muon detectors located in appropriate positions inside the building have been modeled. With a campaign of simulations in different conditions, the performances of the monitoring system in terms of  precision of the measurement of relative displacements and time needed to perform the measurement have been evaluated.  In~\Sref{Other}, these performances are compared with the performances of the mechanical methods adopted in~\cite{Giuriani2000}.

The possibility of constructing a monitoring system as the one described in~\Sref{Muon} strongly depends on the availability of muon detectors with the needed characteristics and performances. In \Sref{Detector}, these requested features are illustrated and a project for the construction of a detector system featuring the requirements is described. Preliminary results of experimental tests of the detector elements are presented. In~\Sref{Summary}, summary and conclusions are drawn.

%
\section{The ``Palazzo della Loggia" of the town of Brescia, monitoring of static anomalies}\label{Loggia}
The ``Palazzo della Loggia" of the town of Brescia, nowadays seat of the Municipal Hall, was completed in 1574 and, since then, it has cumulated a long sequence of injuries, transformations, repairing interventions, some of which have generated considerable problems of structural stability of the building.

The wooden vaulted roof, which is one of the most relevant characteristics of the present aspect of the Palace, was completely reconstructed in 1914 and is representative of a traditional technique used to cover large spans in France and Italy since the seventeenth century.
Its architectural shape and construction technique recall the original roof, destroyed by a fire in 1575, one year after its completion. 

\begin{figure}
\begin{center}
\includegraphics[width=0.7\textwidth]{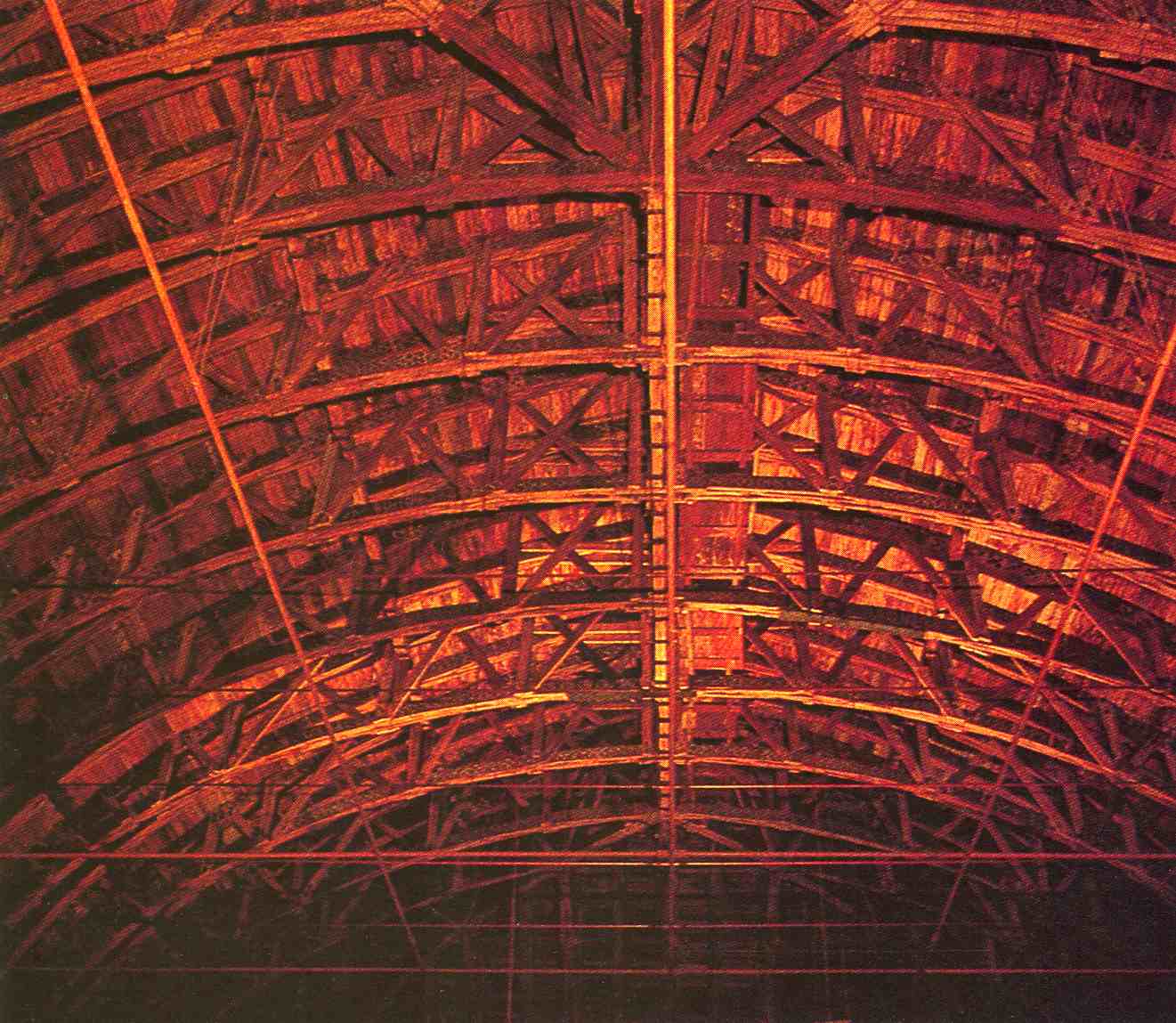}
\caption{\label{Fig2}The present wooden vaulted roof of the ``Palazzo della Loggia", Brescia, Italy. (Courtesy of autors of ref.~\cite{Giuriani2000})}
\end{center}
\end{figure}

The present vaulted wooden structure is of grandiose dimensions and reaches in elevation a maximum of 16~m. The shape of the dome is like an upside down ship, with planar rectangular sides of about 25~m and 50~m respectively. The structural architecture of the vault consists of principal truss wooden arches and simple secondary arches; both are connected at the top by a truss made wooden beam (see~\fref{Fig2}). 

Immediately after its construction, the wooden vaulted roof structure exhibited a progressive deformation of the longitudinal top beam and of the key points of the connected arches. In particular, the progressive deflection of the top beam was measured to be 190~mm in 1923, 520~mm in 1945, 800~mm in 1980 and is visible on the top of the roof in~\fref{Fig1}.

Since 1990, a systematic campaign of investigation and monitoring of the different stability problems of the ``Palazzo della Loggia"  has been committed by the Brescia Municipality to the {\it ``Centro di studio e ricerca per la conservazione e il recupero dei beni architettonici e ambientali dell'Universit\`a di Brescia"}. In particular, the progressive deformations of the principal arches of the wooden vaulted roof was studied with a measurement system specifically designed and the monitoring has been continuously performed for more than ten years, from 1990 to 2001.

\begin{figure}
\begin{center}
\includegraphics[width=0.7\textwidth]{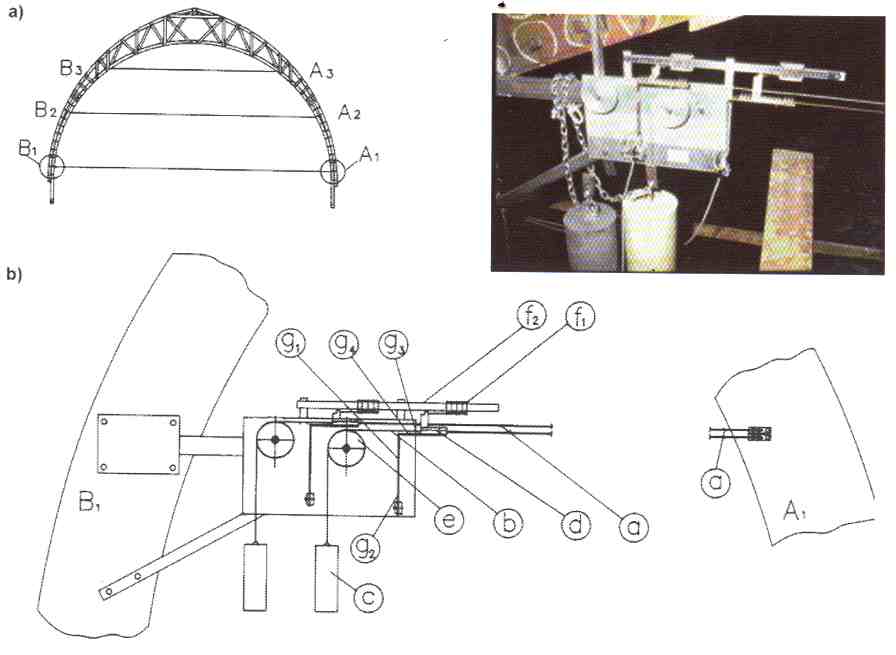}
\caption{\label{Fig3}Monitoring system of the deformation of the truss wooden arches based on couples of stretched wires. Details of the system of pulleys and balance weights and of the electronic and mechanical  measurement systems are shown. (Courtesy of the authors of ref.~\cite{Giuriani2000})}
\end{center}
\end{figure}

The main elements of the measurement system are shown in~\fref{Fig3}. On four, out of the seven principal truss wooden arches, three couples of wires, 2.0~mm  in  diameter, one made of ordinary steel and the other made of invar, are stretched between symmetric points at three different levels: A1 - B1, at the point of connection of the arches with the building structure; A2 - B2 and A3 - B3, on the arch reins.

The wire tension was maintained by means of a system of pulleys and balance weights. The relative displacements of the symmetric points were continuously registered through the differential elongation of the two wires. The different thermal dilatation coefficients of the two different materials made possible to depurate the thermal deformation of the monitoring system itself (see ref.~\cite{Giuriani2000} pag.~193), subject to considerable daily and seasonal thermal variations under the roof covered by lead plates. 

The elongation of the two wires was measured by an electronic system based on clip-gages, with a sensitivity of 1/100~mm, and recorded every six hours. Besides the electronic system, a mechanical measurement system was also employed, based on a vernier with a sensitivity of 1/10~mm, as a check of reliability of the electronic system and recovery for possible failures.

The monitoring system of the wooden vaulted roof remained active for more than ten years. The electronic system was deactivated after three years and  the measurements continued in the following years by means of the mechanical system only. The information obtained was used not only to follow closely the evolution of the progressive deformation of the structure, in order to recognize possible alarm situations, but also to understand the causes of the collapsing and to individuate the most appropriate interventions. 

\begin{figure}
\begin{center}
\includegraphics[width=0.7\textwidth]{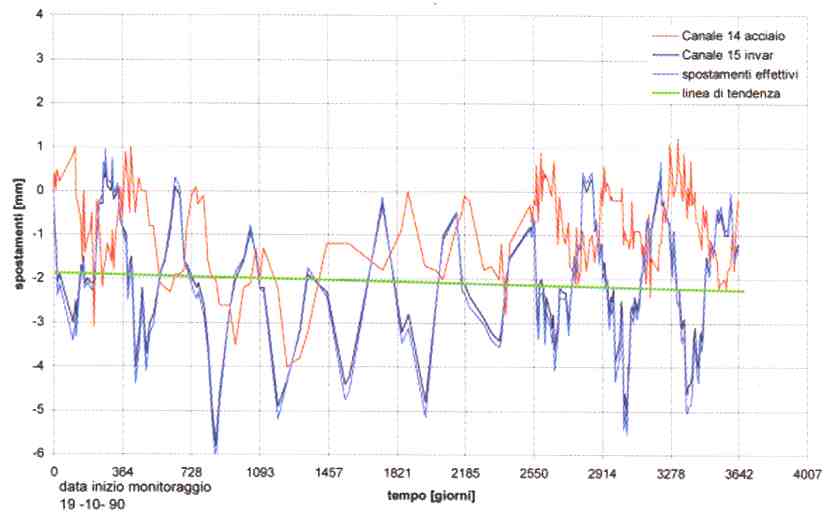}
\caption{\label{Fig4}Elongation of wires stretched between the points A1 - B1 of the truss wooden arch as a function of time in days. Ordinary steel wire (red line), invar wire (blue line), effective deformation of the structure depurated by thermal elongation of the wires (light blue line), general trend of deformation (green line). (Courtesy of the authors of ref.~\cite{Giuriani2000})}
\end{center}
\end{figure}

\begin{figure}
\begin{center}
\includegraphics[width=0.7\textwidth]{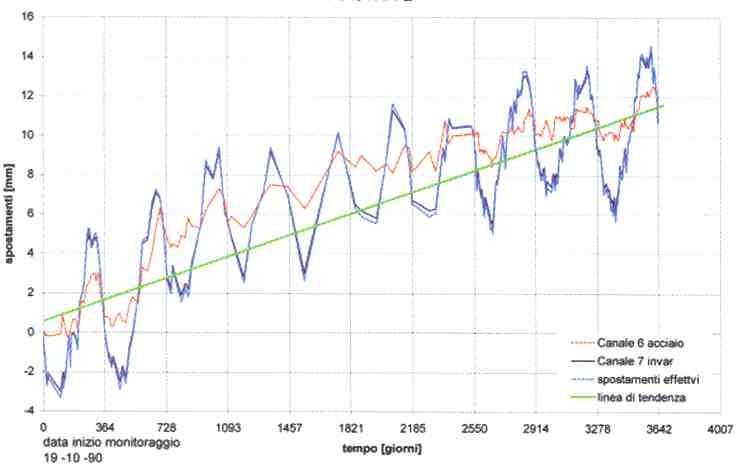}
\caption{\label{Fig5}Elongation of wires stretched between the points A1 - B1 of the truss wooden arch as a function of time in days. Ordinary steel wire (red line), invar wire (blue line), effective deformation of the structure after accounting for the thermal elongation of the wires (light blue line), general trend of deformation (green line). (Courtesy of the authors of ref.~\cite{Giuriani2000})}
\end{center}
\end{figure}

Typical results of the performed measurement campaign are reported in~\fref{Fig4} and \fref{Fig5}. In~\fref{Fig4} the mutual displacements of the points A1 - B1, at the connection of the arches with the building masonry structure, are shown as a function of the monitoring time in days. Elongation of the ordinary steel wire (red line), of the invar wire (blue line) and the effective deformation of the structure after accounting for the thermal elongation of the wires (light blue line), are shown. Effective deformation of the structure practically coincides with the elongation of the invar wire.

Cyclic seasonal deformations of the structure of the order of few millimeters are measured. The general trend (green line) shows, on the contrary, a remarkable stability over the entire monitoring period. The same behavior was found also for the other points of connection of the arches with the building structure, which allowed the deformation of the perimeter supporting masonry walls to be excluded as cause of the wooden vaulted roof deformation.

In~\fref{Fig5} the same data are reported for a couple of wires connecting the arch reins at middle height, in correspondence to the points A2 - B2 in~\fref{Fig3}. Here, superimposed to the seasonal cyclic deformations, a clear collapsing of the wooden structure of the arch is seen. The deformation trend amounts to about 1~mm per year.

%
\section{Application of the muon stability monitoring method to the case of the ``Palazzo della Loggia"}\label{Muon}

\subsection{Assessment of monitoring systems based on the detection of cosmic ray muons}\label{Muon1}
In~\cite{Bodini2007} cosmic ray muon detection techniques were assessed for applications in civil and industrial engineering aimed at the monitoring of alignment and stability of large civil and mechanical structures. Specific reference was made to situations where environmental conditions are weakly controlled and/or where the pieces whose relative positions have to be monitored are hardly accessible.

In that paper, the monitoring of the alignment of a mechanical press, about 5~m high, 1.5~m long and 1.2~m deep, was considered as a representative example. The muon detection system was composed of a telescope of three generic muon hodoscopes, in form of square sensitive plates, 200~mm side and 10~mm thick, made of plastic organic scintillator, a typical constituent of common particle detectors.

The intrinsic position resolution of the muon detectors along both coordinates of the sensitive surface was considered to be $\sigma$=100~$\mu$m. The three muon hodoscopes were positioned perfectly aligned along a vertical axis and mechanically connected to the parts of the structure that have to be monitored, in the upper, middle and lower parts of the mechanical press. The distance between the upper and lower detectors was 3.3~m and the distance between the upper and the middle detectors was 2.5~m. 
The mechanical structure of the press was considered made of iron. Between the middle and the lower detectors, a structural part of the press, an iron layer of 280~mm thickness, was positioned.

The described set-up was modeled in a Monte Carlo simulation based on the GEANT4 toolkit for the simulation of the passage of particles through matter~\cite{Agostinelli2003}, taking into account the geometry of the press structure and the detector telescope. A realistic cosmic ray generator based on experimental data was implemented in the code in order to simulate, as realistically as possible, the momentum, angular distributions and other features of the cosmic ray radiation at the Earth surface.

Extensive Monte Carlo simulations of the tracking of cosmic rays through the simulated set-up allowed the expected measurement uncertainty for the monitoring of the mechanical press alignment to be assessed. In addition, its dependence on the system geometry, presence of interposed materials between the detectors, intrinsic resolution of the detectors and elapsed time available for the measurement was determined.
 
In particular, the position of the middle detector (relative to the other two) was determined by recording the hitting positions of cosmic ray muons crossing the whole telescope and performing an appropriate statistical analysis of the simulated data.
In a fixed geometrical configuration, the standard uncertainty on the measurement of the possible misalignment of the middle detector is mainly connected to the number of events of crossing muons registered by the telescope, which is linearly proportional to the time available for the data taking.

As an example of the obtained results, supposing a calibration time of one week, necessary to establish the reference position of the middle detector with a measurement standard uncertainty of about 85~$\mu$m, the system should be able to detect a relative displacement of 1.0~mm in about 12~h of data taking, whereas 60~h are needed to detect a 0.5~mm displacement at 99.85\% level of significance. 

Factors affecting the system resolution performances are the telescope geometry (detector sensitive surface and detector distances), which is connected with the solid angle for cosmic ray acceptance, and the thickness and position of layers of materials interposed among the detectors, which determines the amount of deviations due to multiple scattering of the muons crossing the telescope.
Detector intrinsic spatial resolution, on the contrary, does not seem to be a crucial parameter. 

This method could be particularly useful when the parts that have to be monitored are not reciprocally visible. In addition, it makes use of a natural form of radiation, ubiquitously present on the entire Earth surface, without  need of artificial sources and radiological protection means. The low cosmic ray rate and its wide angular distribution around the zenith direction, which implies long data taking time to obtain significant statistics, is the main limiting factor for the application of the described method.

%
\subsection{The case of the wooden vaulted roof of the ``Palazzo della Loggia"}\label{Muon2}
In~\cite{Bodini2007} it was suggested that muon stability monitoring technique could be effectively applied also to civil engineering, to supervise static stability of dams, historical monuments, such as towers or belfries, and buildings in seismic areas. The advantages of the method are low upkeep and steady data acquisition, with reachable uncertainties comparable to those affecting other monitoring systems typically used in this field.

In the present paper, this suggestion is applied to the case of the ``Palazzo della Loggia" of the town of Brescia, on which the highly penetrating cosmic ray muons rain continuously at a rate of 210,000~$\mu$/s, in order to evaluate the potential performances of the method compared with the results obtained with the measurement procedure actually applied.

The problem of the monitoring of the wooden vaulted roof of the ``Palazzo della Loggia", as described in~\Sref{Loggia}, can be synthesized as follows: the possible displacements of a certain number of constituent elements of the wooden vaulted roof must be monitored relative to a reference coordinate system fixed with the masonry structure of the building. 
The total elapsed time over which the monitoring should be performed is of the order of years, and the frequency of sampling measurements of the order of days or weeks.

It is worth remarking that the mechanical method actually adopted in~\cite{Giuriani2000} could only provide the measurement of the horizontal relative displacement of points of the wooden structure in opposite positions and not their absolute displacements relative to a common reference system linked to the building structure. Some remarks on the possible use and expected performances of different standard monitoring methods are given in next~\Sref{Other}.

\begin{figure}
\begin{center}
\includegraphics[width=0.4\textwidth]{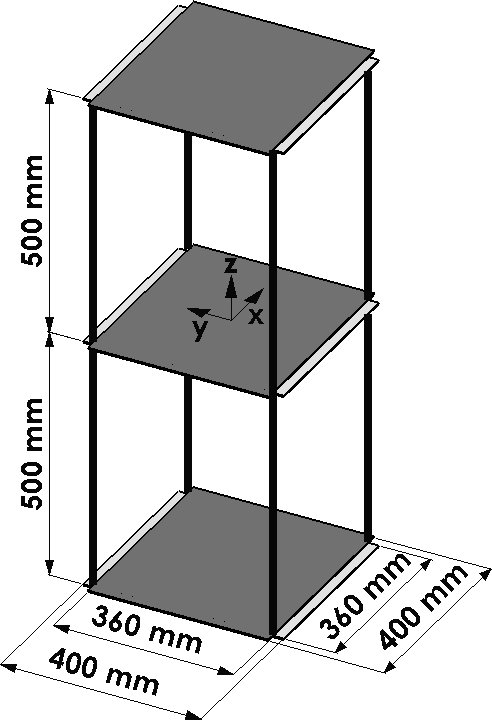}
\includegraphics[width=0.4\textwidth]{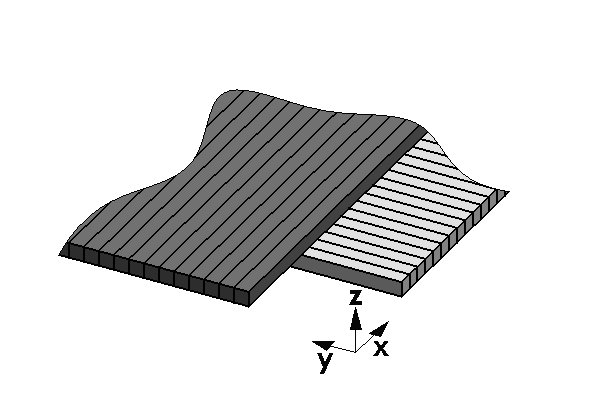}
\caption{\label{Fig6}a) Structure of the ``muon telescope" formed of three muon hodoscope modules axially aligned at a distance of about 50~cm each other. b) Sensitive volume of the muon hodoscope module formed of two orthogonal layers of 120 scintillating fibers 3.0~mm~$\times$~3.0~mm cross section and 400~mm length. }
\end{center}
\end{figure}

A monitoring system based on cosmic ray detection, able to achieve the task described above, has been designed in its general geometrical structure as follows. A set of three square muon hodoscope modules, of about 400~mm side and 6.0~mm thickness, are positioned on an appropriate mechanical structure and axially aligned at a distance of 50~cm one from the other. This set-up corresponds to a ``muon telescope" of total length of about 100~cm, as shown schematically in~\fref{Fig6}a). The ``muon telescope" is composed of three modules to allow for a minimum level of redundancy on the tracking information for the crossing muon. 

The geometry of the sensitive volume of each hodoscope module is formed of two orthogonal layers, 360~mm width, composed of 120 scintillating fibers with 3.0~mm~$\times$~3.0~mm cross section and 400~mm length. The two layers are arranged along two $x$ and $y$ axes of a Cartesian reference system as shown in~\fref{Fig6}b). 
The two layers of orthogonal square scintillating fibers provide the measurement of the crossing position of an incident muon in the $x$ and $y$ coordinates with a pitch of 3.0~mm. Considering a flat detection efficiency over the full surface of the scintillating fiber, the expected spatial resolution on the hit coordinate should be $\sigma$=(3.0~mm)/$\sqrt{12}$=0.87~mm. 

\begin{figure}
\begin{center}
\includegraphics[width=0.6\textwidth]{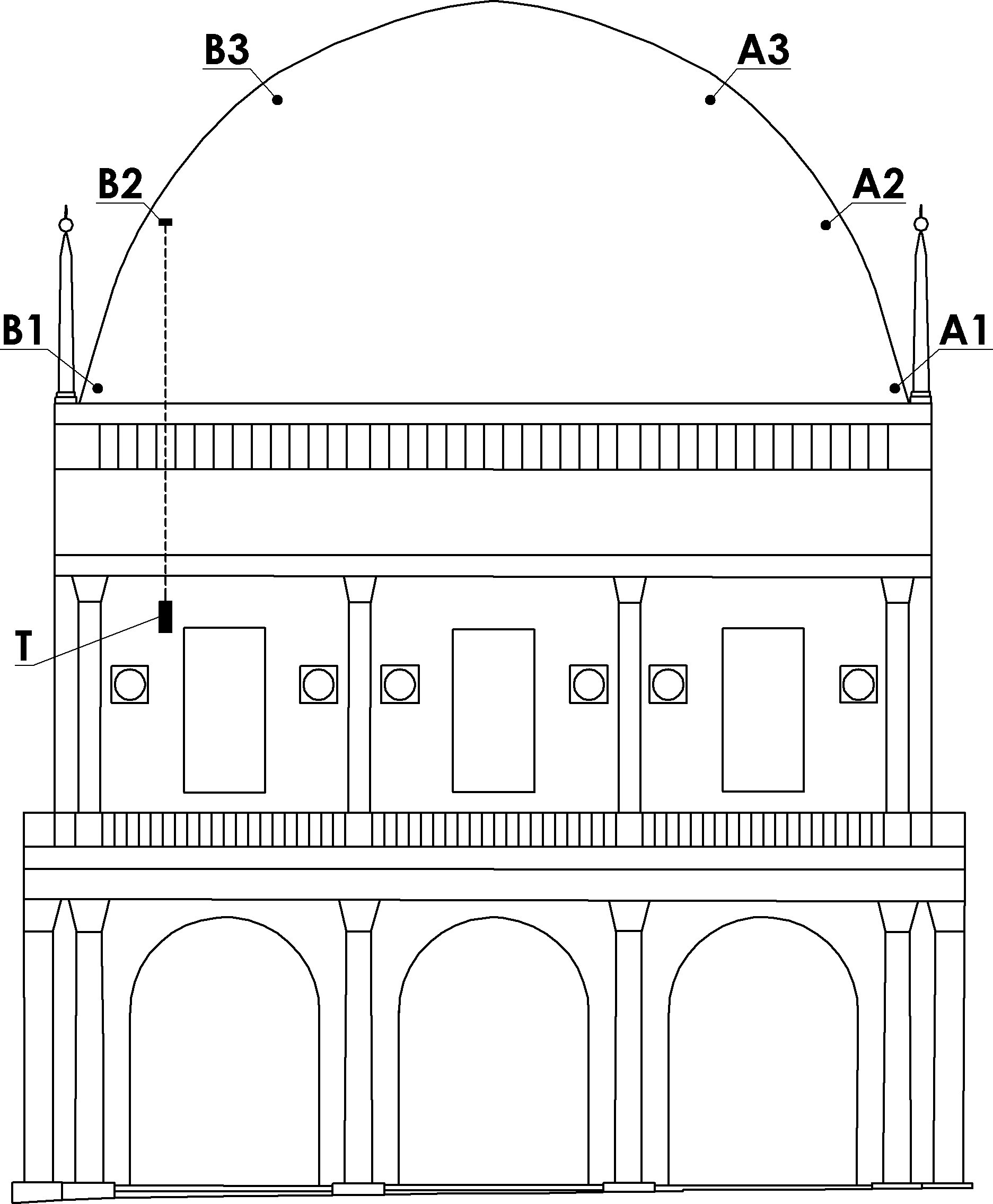}
\caption{\label{Fig7}Schematic cross section of the ``Palazzo della Loggia" building with superimposed the ``muon telescope" and the ``muon target" positioned respectively on a structural element of the building (the fixed reference system) and on the point of the building to be monitored.}
\end{center}
\end{figure}

As shown in \fref{Fig7}, the ``muon telescope" must be mechanically fixed to a structural element of the building, which is considered as the reference system for the measurement of the position of the building points to be monitored. The axis of the telescope should be aligned in the direction corresponding to one of these points: in the case considered, points B1, B2 or B3 of the wooden arches of the roof. A fourth muon hodoscope module, with the same geometry and structure of the previous ones, must be positioned as ``muon target" on the building element which has to be monitored, as much as possible coaxial with the ``muon telescope". 

Cosmic ray muons traversing the full system of four detectors and the interposed structures of the building may allow the displacement of the ``muon target" relative to the fixed ``muon telescope" to be monitored continuously. The precision of the position measurement depends on the system geometry, the interposed materials, the data taking time. Recognition of signals generated by the same cosmic ray muon traversing the four detectors, against background casual events generated by different cosmic ray muons, may be efficiently insured by time coincidence of the four detector signals~\cite{ALICE2010}.

Indeed, the direction and crossing point of any cosmic ray traversing the ``muon telescope" can be measured with precision depending on the telescope geometry and detector granularity. The measured cosmic ray trajectory can be extrapolated from the ``muon telescope" to the plane of the ``muon target" detector, in the hypothesis that the trajectory of the muon be a perfect straight line. The distance between the ``muon telescope" and the ``muon target" can be deduced, with sufficient approximation, from the building drawings. Due to the geometry of the system, which limits the solid angle covered by cosmic ray muons traversing all the four detectors, uncertainties on the determination of this distance has negligible effects on the determination of the extrapolated crossing point. 

The coordinates of the extrapolated crossing point can be compared with the coordinates of the crossing point of the same cosmic ray measured with the ``muon target" module. From the differences of these two sets of coordinates, possible displacements of the position of the ``muon target" module relative to a reference position previously determined can be inferred. 

The hypothesis that the muon trajectory is a perfect straight line is justified only in the absence of any magnetic field and in the vacuum. The interaction between the Earth's magnetic field and, for example, a 3.0~GeV/c muon causes a deflection of about 5~$\mu$m/m in the direction perpendicular to both magnetic field and particle velocity. Since in cosmic ray flux there is an excess of positive muons with respect to negative ones, this could induce a systematic effect on the measurement. However, the proposed method is based on a relative measurement, say considering differences between measurements performed at different times, thus the effect of the terrestrial magnetic field can be neglected. 

More relevant is, of course, the effect of materials interposed on the muon paths, which determines stochastic deviations of the muon trajectories due to multiple scattering with atomic nuclei~\cite{Moliere1947,Moliere1948,Bethe1953}. The angular deviation variance is proportional to the density of the material, its atomic number and the total amount of crossed material, and inversely proportional to the square of the muon momentum.

The uncertainty in the prediction of the crossing point coordinates of the muon, in the ``muon target" plane, starting from the measurement of its trajectory in the ``muon telescope", will depend on: position resolution of the muon detectors; geometry of the ``muon telescope"; distance of the ``muon target" from the ``muon telescope"; amount of multiple scattering angular deviations and displacement of the muon trajectory, which depend on the amount of interposed materials and their positions. 

Being these stochastic effects largely dominant, statistical distributions of the difference between measured crossing point coordinates in the ``muon target" and the predicted crossing point coordinates obtained by extrapolation from the ``muon telescope" is therefore necessary, in order to reduce the stochastic effects by statistical inference methods.  Using the methods developed in~\cite{Bodini2007}, efficient unbiased estimators of the systematic displacement can be extracted by means of a statistical analysis of the distributions. In next subsections the features and expected performance of such a measurement system will be calculated by a Monte Carlo simulation.

Finally, it is worth remarking that, depending on the solid angle covered by the ``muon telescope" and on the position and distances of the points to be monitored, more than one ``muon target" could be monitored simultaneously by the same ``muon telescope", with only some reduction of the acceptance of the system for the points more out of axis. In this way, a global and simultaneous stability monitoring  of several parts of the building can be performed.

%
\subsection{Simulation of the muon stability monitoring system in the ``Palazzo della Loggia"}\label{Muon3}
The simulation was performed by GEANT4 \cite{Agostinelli2003}, a C++ toolkit for the simulation of the passage of particles through matter largely utilized in the design of nuclear and particle physics experiments. The geometry and relevant structural parts of the ``Palazzo della Loggia" building were taken into account, the structure and composing materials of the ``muon telescope" and ``muon target" were modeled. 

In three separated simulations, the ``muon telescope" was located in three different positions, 3.0~m below the ceiling of the large ``Salone Vanvitelliano" at the first floor of the Palace, as shown in \fref{Fig7}, on the vertical of each one of the three points of the wooden vaulted roof to be monitored: B1, B2 and B3 of \fref{Fig3}. These points are positioned, respectively, 0.50~m, 5.8~m and 10.0~m above the ceiling of the ``Salone Vanvitelliano". In these three points  a ``muon target" was located for each of three separated simulations, on the vertical of the corresponding ``muon telescope".

The ceiling of the ``Salone Vanvitelliano" was modeled as a bulky wooden layer, 15.0~cm thick. No vertical structures of the Palace were modeled, in order to study with the simulation the intrinsic limits of the monitoring system, nor the wooden vaulted roof covered by lead plates, since it would have introduced only a negligible distortion of the incoming cosmic ray spectrum and angular distribution, without sensitive effects on the system performances.

Finally, a realistic cosmic ray muon generator based on experimental data was  implemented in the code in order to simulate, as realistically as possible, the momentum, the angular distribution and the charge composition of the cosmic ray radiation at the sea level \cite{Bonechi2005}. The cosmic ray muon momentum distribution is almost flat for momenta below 1~GeV/c and falls as p$^{-2.7}$ for momenta above 10~GeV/c. The mean muon momentum is about 3-4~GeV. The flux is greatest at the zenith and falls approximately as $\cos ^2 \theta$, where $\theta$ is the plane angle from the vertical. The overall muon rate is about 10,000~$\mu$/(min~m$^2$) for horizontal detectors.

\begin{figure}
\begin{center}
\includegraphics[width=0.7\textwidth]{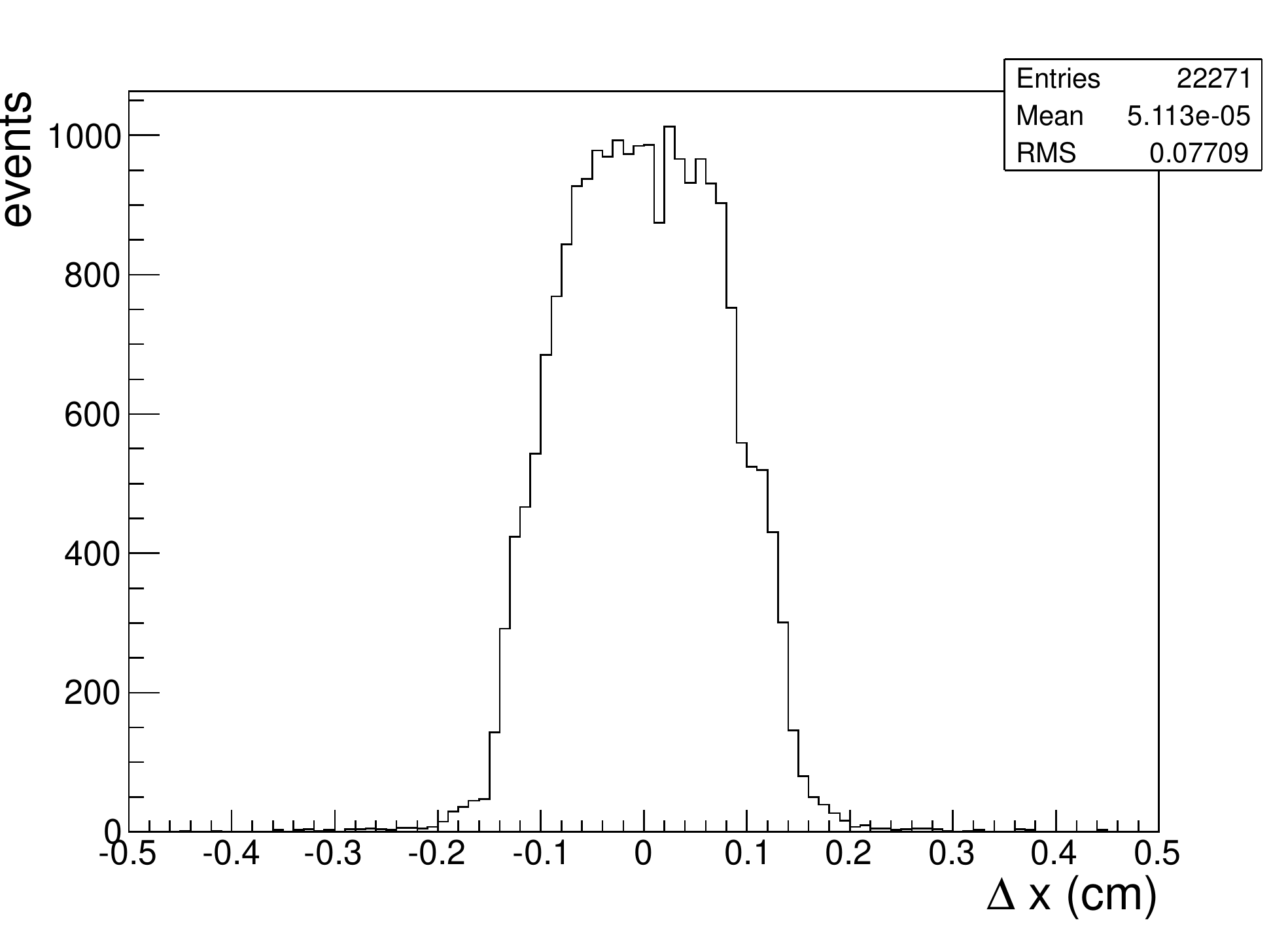}
\caption{\label{Fig8}Distribution of the difference between the $x$ coordinate of the point of generation of the cosmic ray on the surface of the upper detector module of the ``muon telescope" and the same coordinate reconstructed by the straight line fit.}
\end{center}
\end{figure}

A preliminary simulation was devoted to the evaluation of the resolution of the ``muon telescope" in the measurement of the direction of the cosmic ray crossing the telescope and of the coordinates of the crossing point on the plane of the upper detector module. To this aim, a dedicated simulation has been performed generating a population of cosmic ray muons originating randomly on the surface of the upper detector module.

In each detector module the passage of the muon is registered when an amount of energy is released by ionization energy loss in one scintillating fiber; the measured position of the muon crossing point in $x$ and $y$ coordinates is defined by the position of the axis of the hit scintillating fiber, on the corresponding layer, in the coordinate system shown in~\fref{Fig6}(a). 

To simplify the analysis, only the muons providing one and only one hit for each of the three scintillating fiber layers on the same coordinate are considered. For this sample of cosmic ray muons, the three measured points on the three layers for each coordinates are fitted with a straight line. The direction of the reconstructed straight line is compared with the direction of the generated muon at the entrance of the ``muon telescope" and the $x$ and $y$ coordinates of the crossing point of the straight line with the upper surface of the upper detector module is compared with the coordinates of the muon origin point randomly extracted. 

\begin{figure}
\begin{center}
\includegraphics[width=0.7\textwidth]{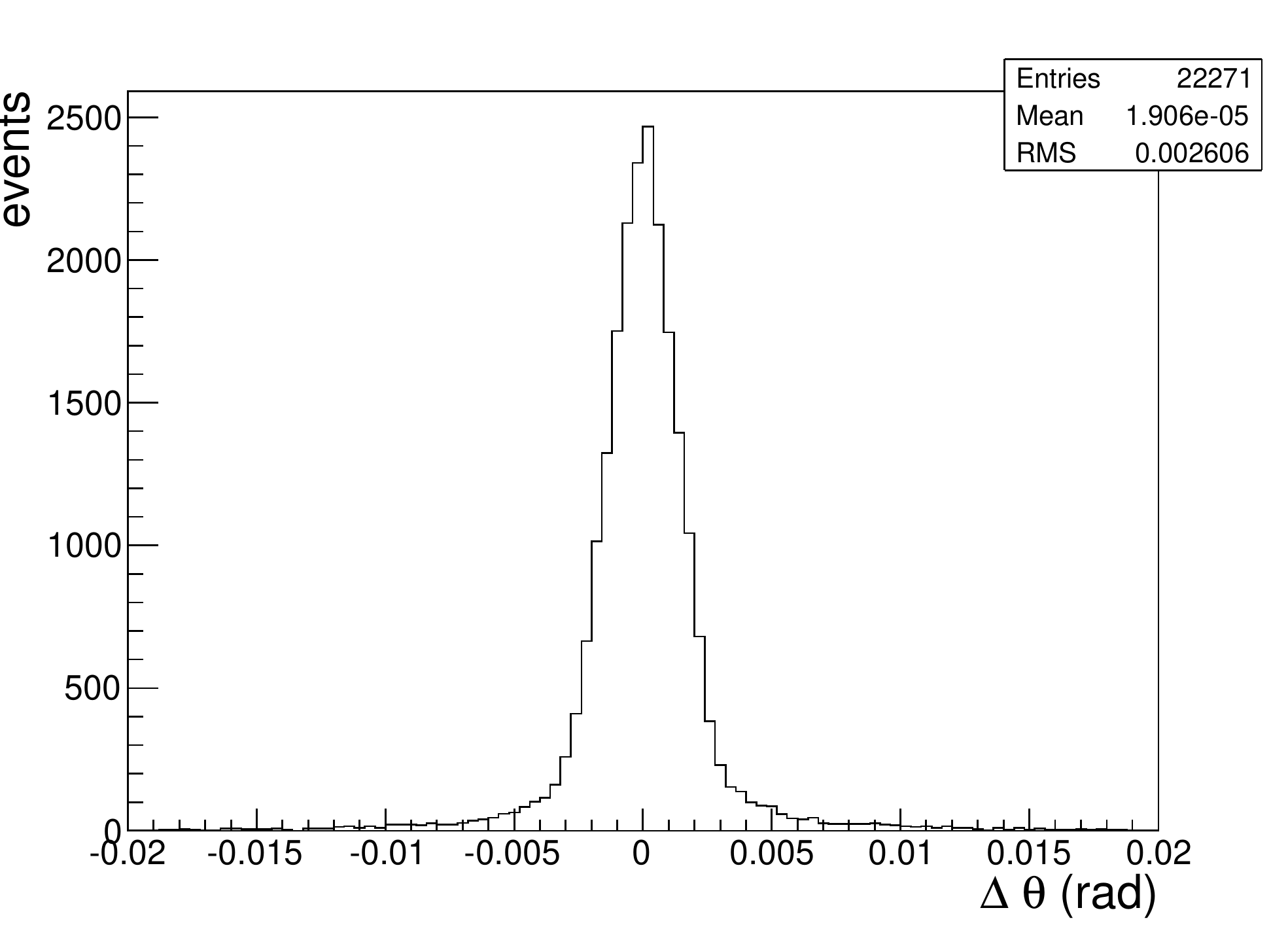}
\caption{\label{Fig9}Distribution of the difference between the projected angle on the $x$-$z$ plane of the generated cosmic ray crossing the ``muon telescope" and the same angle reconstructed with the straight line fit.}
\end{center}
\end{figure}

In~\fref{Fig8} the distribution of the difference of the $x$ coordinates of the generated point and of the reconstructed point is shown; the system being symmetrical, the distribution for the $y$ coordinate is the same from a statistical point of view. The distribution is obtained with a statistics corresponding to a data taking time of 6~hours. The standard uncertainty in the determination of the muon crossing point is about 0.8~mm, compatible with the intrinsic granularity of the muon detector modules.

In~\fref{Fig9} the distribution of the difference between the projected angle on the $x$-$z$ plane of the generated cosmic ray and the reconstructed one is shown. The standard uncertainty in the determination of the muon projected direction is 2.6~mrad, compatible with the granularity and the geometry of the telescope, which suggests an uncertainty of about 3~mrad. 
With such a precision in the determination of the direction of the cosmic ray muon, the expected contribution of the ``muon telescope" uncertainty in the determination of the muon crossing point on the ``muon target" plane positioned at 10~m distance is about 3.0~cm, say 0.3~cm every meter in distance. 
As it will be seen in next subsection, the contribution to the uncertainty given by the multiple scattering deviations of the muon trajectories is largely dominant in this configuration.

%
\subsection{Position measurement uncertainty of the muon stability monitoring system versus data taking time}\label{Muon4}
With the described set-up, simulations of tracking of cosmic ray muons through the measurement system were performed in the three configurations described above. For cosmic ray muons giving one and only one hit in the four detector modules, the differences $\Delta x$ and $\Delta y$  between the crossing point coordinates measured in the ``muon target" and the crossing point coordinates of the extrapolated muon trajectory measured by the ``muon telescope" are recorded. All relevant physical processed in the generation and tracking of the cosmic ray muons are taken into account in the simulation program. 

Cosmic ray muons generating multiple hits in the same detector layer are rejected from the analysis. Multiple hits may be produced by the crossing of more than one scintillating fibers in one layer or by showering of the cosmic rays muons and possible generation of delta rays. The number of cosmic ray muons producing multiple hits in the same detector is about 15\% of the total number entering the measurement system acceptance. A more refined data analysis and pattern recognition may allow a part of these events to be recovered for analysis.

\begin{figure}
\begin{center}
\includegraphics[width=0.48\textwidth]{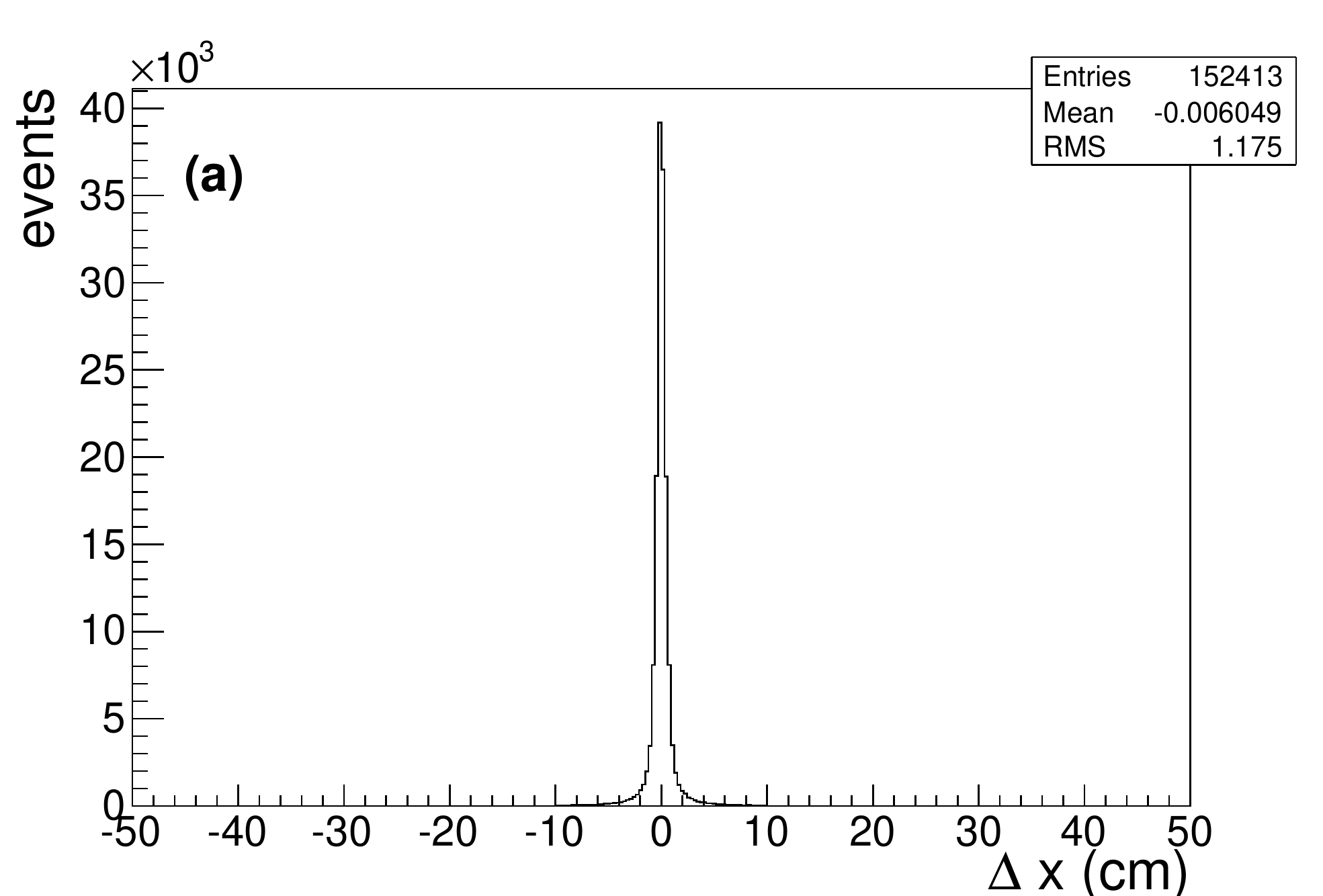}
\includegraphics[width=0.5\textwidth]{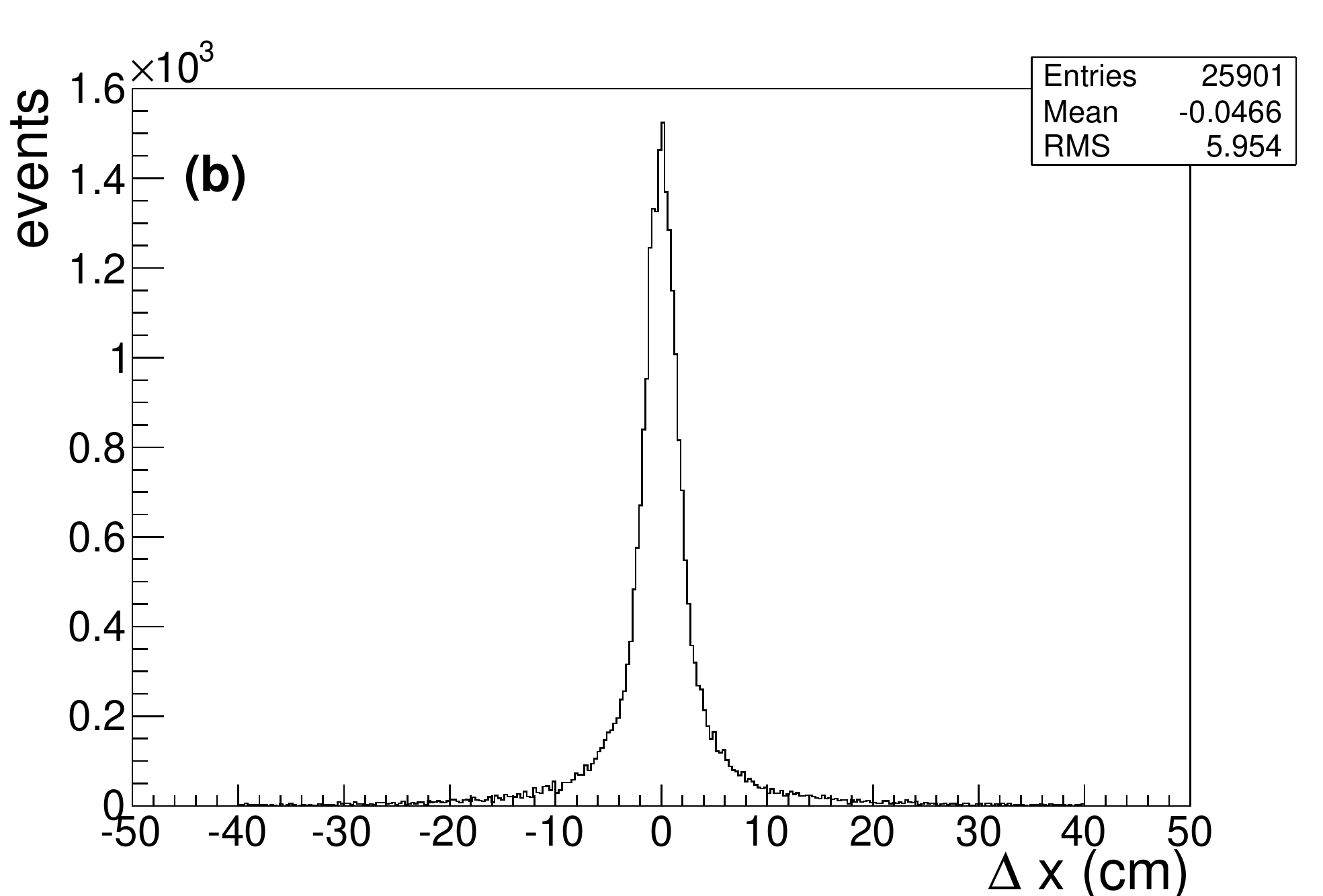}
\includegraphics[width=0.5\textwidth]{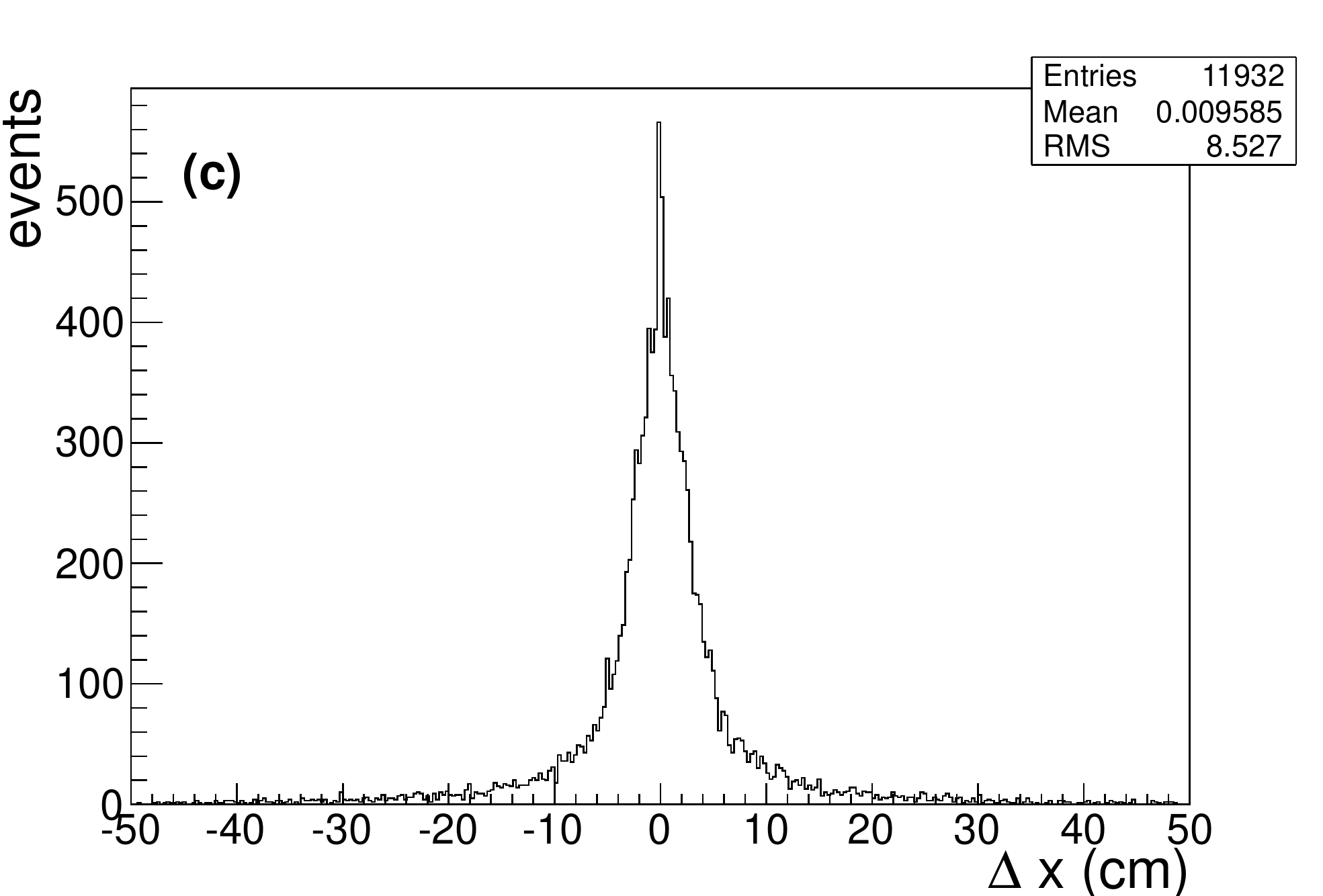}
\caption{\label{Fig10}Distributions of the difference $\Delta x$  between the crossing point coordinates measured on the ``muon target" in position B1 (a), B2 (b) and B3 (c), and the crossing point coordinates of the extrapolated muon trajectory measured by the ``muon telescope". Simulated data taking time is 15 days.}
\end{center}
\end{figure}

In~\fref{Fig10} the distributions of the statistical variable $\Delta x$ are shown for the three configurations of the measurement system described above, for an elapsed data taking time of 15 days, corresponding to about 31.7$\cdot 10^6$ cosmic ray muons crossing the ``muon target" surface. The $\Delta y$ distributions are not shown, since they are statistically identical to the $\Delta x$ ones.
The number of events in each distribution is compatible with the number of expected cosmic ray muons entering the geometrical acceptance of the measurement system given by:
\begin{equation}
\label{eq1}
N_{acc} = (\frac{d^2 N}{d \Omega_{xy} dA_1}) \cdot (\frac{A_2}{R^2}) \cdot A_1 \cdot T 
\end{equation}
where $(\frac{d^2 N}{d \Omega_{xy} dA_1})$=70~$\mu$/{(m$^2$~s~sr)} is the muon rate per unit area and unit solid angle around the zenith direction~\cite{Beringer2012}, $A_1$ is the surface of the ``muon target", $A_2$ is the lowest surface of the  ``muon telescope" and $R$ is their distance. $T$ is the elapsed time of the data taking in seconds.

As the ``muon target" and the ``muon telescope" are exactly coaxial in the simulation, the $\Delta x$ distributions are symmetric and centered at zero. The shape of the distributions exhibits a central narrow peak with very long tails on both sides. This shape is due both to the intrinsic uncertainty of the ``muon telescope" in measuring the direction of the cosmic ray muon and to the multiple scattering angular deviations of the muon trajectories traversing the interposed materials. 
At fixed momentum and little angles, these deviations follow a Gaussian law with variance depending on the inverse square of the muon momentum~\cite{Bethe1953}. 

The latter effect dominates for larger distances of the ``muon target" from the ``muon telescope".  The long tails of the distributions are due in part to low momentum muons, suffering larger deviations, and, in part, to spurious events corresponding to emission of delta rays or cosmic ray showering, most of which can be discarded with a more refined data analysis. At present, the only selection applied to these bad quality events is an arbitrary cut of both tails in the three distributions, discarding about 1.0\% of the total events.

The root mean squares of the $\Delta x$ distributions are 1.18~cm, 5.95~cm  and 8.53~cm respectively for the three configurations examined. They represent the uncertainties in the prediction of the crossing coordinate of the cosmic ray muon on the ``muon target" detector extrapolated by the muon trajectory measured with the ``muon telescope". 

This uncertainty depends mainly on the distance between ``muon target" and ``muon telescope" and on the amount and position of interposed materials. The mean value of the sample distribution represents an unbiased estimator of the position of the ``muon target" relative to the ``muon telescope" axis. The uncertainty on the mean value of a sample distribution is given by the well known relation~\cite{Rotondi2001}:
\begin{equation}
\label{eq2}
\sigma_{mean} = \sigma_{distr}/\sqrt {N_{ev}} 
\end{equation}
where $N_{ev}$ is number of events in the distribution. In the three configurations considered and for a data taking time of 15~days, these standard uncertainties are respectively 0.03~mm, 0.37~mm, 0.78~mm. 

\begin{figure}
\begin{center}
\includegraphics[width=0.7\textwidth]{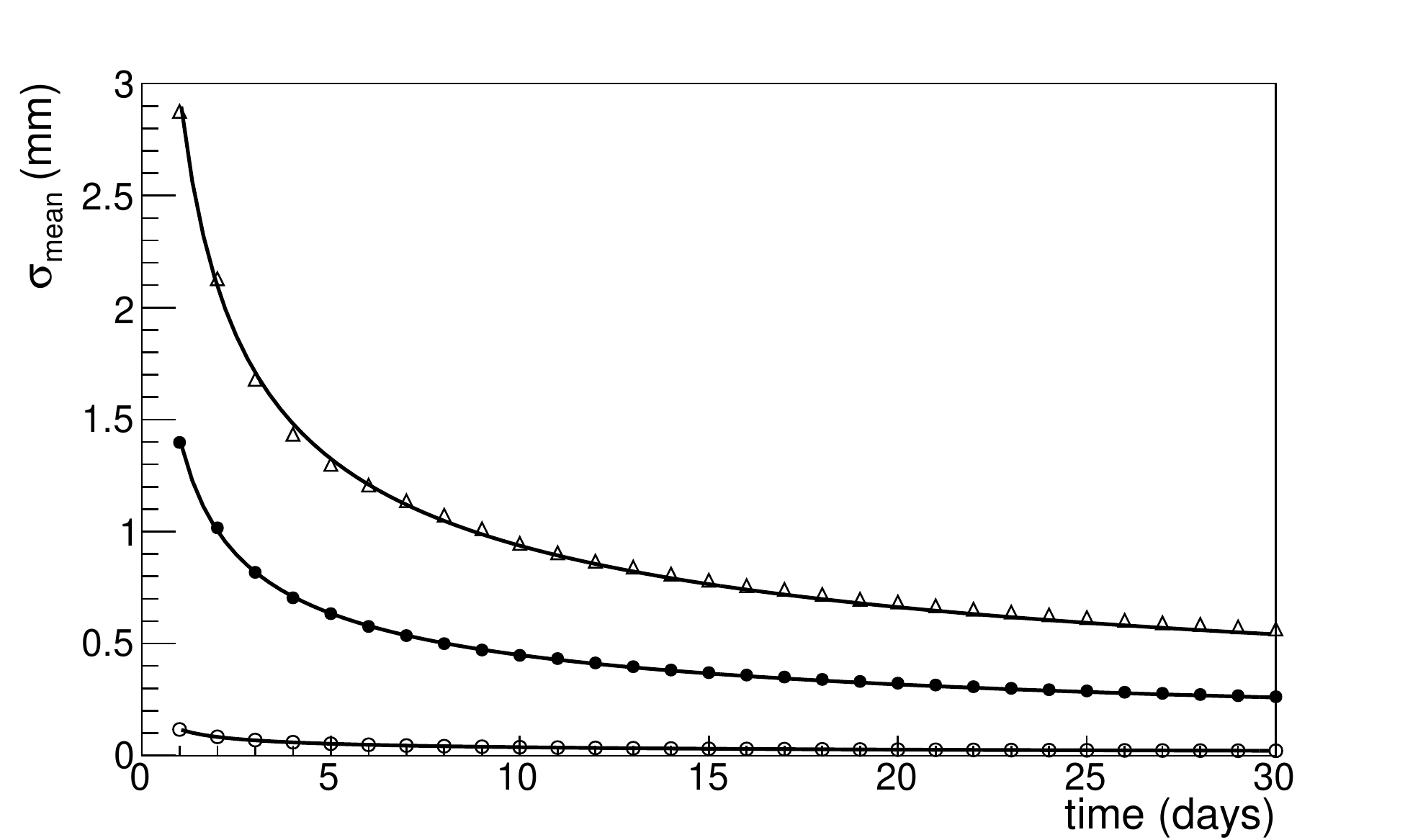}
\caption{\label{Fig11}Standard uncertainty on the mean value of the sample distributions versus data taking time, for the ``muon target" in position B1 (\opencircle), B2 (\fullsquare) and B3 (\opentriangle).}
\end{center}
\end{figure}

Since in the same geometrical condition $N_{ev}$ is simply proportional to the data taking time, the measurement standard uncertainty depends only on the inverse of the square root of the data taking time.  In~\fref{Fig11} the relation of the position measurement standard uncertainty and the data taking time for the three examined conditions is plotted up to a data taking time of one month. As time increases the measurement standard uncertainty decreases. By fitting the plots with the general relation \cite{Bodini2007}:
\begin{equation}
\label{eq3}
\sigma_{mean} = C /\sqrt {t} 
\end{equation}
where $C$ is a constant depending on the geometry and materials interposed and $t$ is the data taking time in days, the following values for the constant $C$ are obtained in the three conditions considered: 1.12~mm~day$^{1/2}$, 1.42~mm~day$^{1/2}$, 2.96~mm~day$^{1/2}$, with the standard errors of the order of 1\%.

As expected, the standard uncertainty of the measurement system depends on the geometrical configuration considered, since both the root mean square of the $\Delta x$ distributions and the rate of useful events collected are strongly dependent on the geometry of the system and on the amount of materials interposed.

Nevertheless, although requesting different data taking times, the monitoring of the displacement of the three inspected points in the wooden vaulted roof of the ``Palazzo della Loggia", using a cosmic ray tracking system, could provide performances compatible with the requested precisions and with the time scale characteristic of the  deformation phenomenon. Typical time scales, in the case of ``Palazzo della Loggia" and, in general, for historical buildings, may span over several years. 

In position B1, a measurement standard uncertainty of the order of 0.1~mm may be achieved in about one day of data taking, whereas a standard uncertainty of the order of 0.5~mm may be achieved in a week of data taking in position B2, and in one month of data taking in position B3, where ``muon target" and ``muon telescope" are positioned 13.0~m far apart.

%
\subsection{Possible improvements of the performances of the muon stability monitoring system}\label{Muon5}
It is worth pointing out that the performances of the described measurement system can be enhanced in two different ways, both acting on system parameters and by improving the data analysis. Indeed, equation~\eref{eq1} shows that the number of cosmic ray muons entering the acceptance of the measurement system is proportional to the product of the cross section surfaces of the ``muon target" and the ``muon telescope".
If the detector modules have the same effective surface, as in the case considered, an increase of 50\% of the single module size may improve by a factor 5 the number of accepted cosmic rays, reducing of a factor 2.2 the data taking time needed to obtain the same standard uncertainty on the position measurement.

Further improvement can be obtained, as pointed out in~\cite{Bodini2007}, discarding from the muon sample the low momentum muons by means of a muon adsorber in form of a thick iron plane positioned below the ``muon telescope" and followed by a further plane scintillation counter in coincidence. Eliminating the low momentum muons, which suffer the largest multiple scattering deviations, and the possible electron showering that adds spurious events to the sample, the long tails of the  $\Delta x$ and  $\Delta y$  distribution can be reduced, reducing consequently the root mean square $\sigma_{distr}$ of the distributions.

As concerns the improvements in the data analysis, it was demonstrated in~\cite{Bodini2007} that a more efficient unbiased estimator of the position and possible displacement of the ``muon target" detector can be obtained by fitting the parent population of the statistical variable  $\Delta x$ by an appropriate analytical function and determining its shape parameters by the best fit procedure. 
The same best fit function, with only two free parameters, the first one representing the displacement of the whole function shape along the $x$ coordinate (typically a parameter representing a central value or the mean value of the distribution) and a second one giving the global scale factor, is then fitted to the distribution of a sample of events.

The position parameter provides an estimator of the same parameter of the parent population, which is more efficient than the sample mean value is for estimating the population true mean value. This result appears reasonable, since the use of an analytical function approximating the shape of the parent population adds to the estimation procedure extra information coming from the full shape of the population, in comparison with the simple mean value parameter of the sample. The standard uncertainty of the position parameter of the best fit function is provided, in a reliable way, by the best fit procedure.

In the case studied in~\cite{Bodini2007}, the shape of the parent population for the target detector was represented using a fitting function obtained as the sum of a number of Gaussian functions with the same mean and different standard deviations and weights. The reason for this choice relays on the fact that, for a fixed value of the muon momentum, the physical process of multiple scattering produces, at a first approximation, a Gaussian distribution of the deviation angle. 

However, the spectrum of the cosmic ray muons at ground level is rather wide and the material thickness that they have to cross may be different from a cosmic ray to another one. In symmetric geometrical conditions, a sum of Gaussian functions resulted therefore a good choice for representing the parent population. 

It is worth remarking that the parent population of the statistical variables under study may be obtained in two ways. Either experimentally, when the conditions of the considered system make it possible, by a long calibration data taking, or, alternatively, by a Monte Carlo simulation of the process, taking into account in detail the structure and materials crossed by cosmic ray muons in all their possible flight paths through the system. 

In~\cite{Bodini2007} it was shown that the estimator of the true mean value of the parent population obtained by the fitting function procedure described above is 2.5 to 3.0 times more efficient than the simple sample mean value. This is indeed the ratio of the standard uncertainties of the two estimators on the same samples of data.

\begin{figure}
\begin{center}
\includegraphics[width=0.7\textwidth]{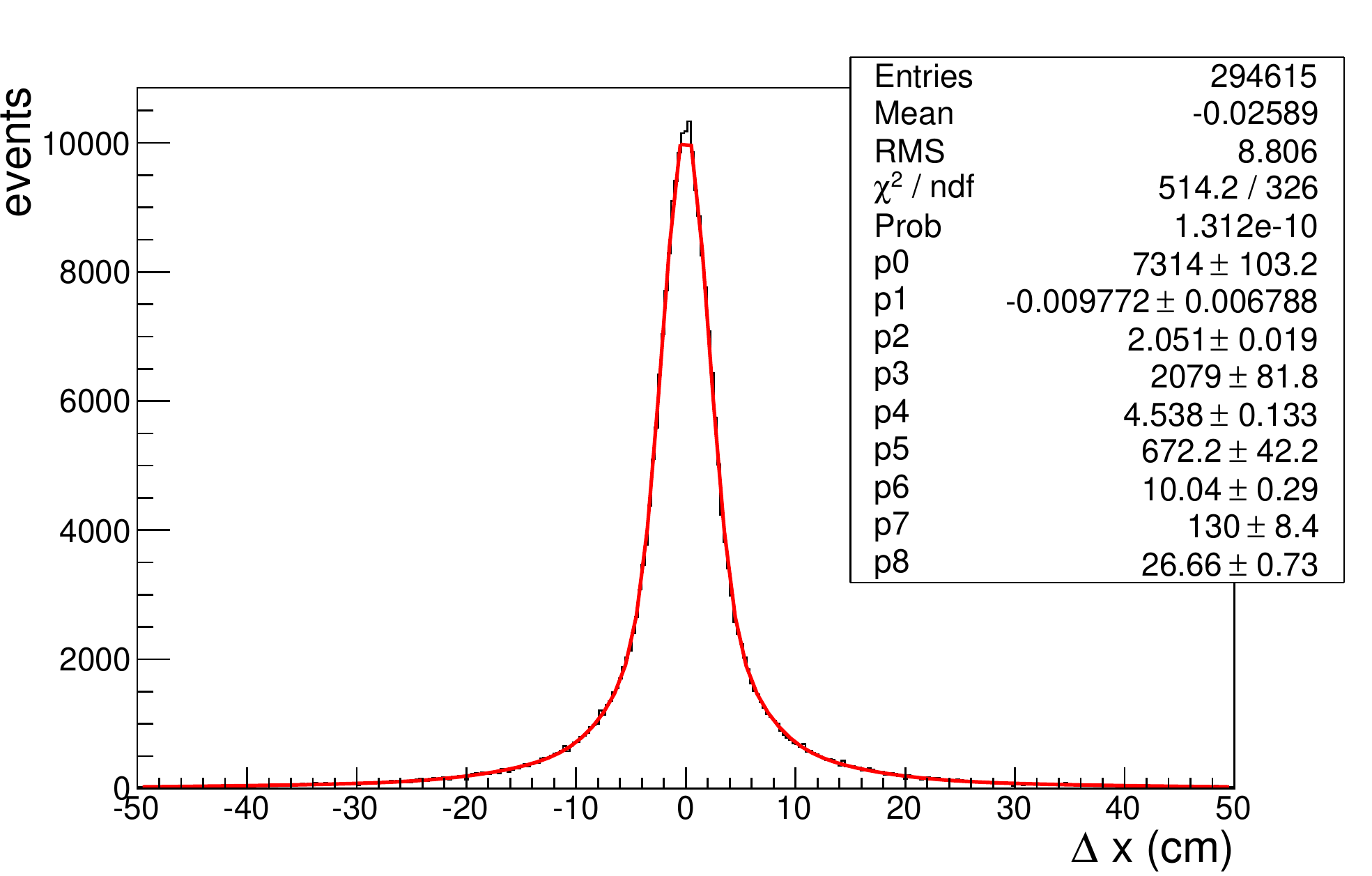}
\caption{\label{Fig12}Distribution of the statistical variable  $\Delta x$ for the ``muon target" in position B3 corresponding of a data taking of one year. The distribution likely represents the variable population. It has been fitted by a fitting function composed of the sum of four Gaussian functions with the same mean (p1) and different standard deviations (p2, p4, p6 and p8) and weights (p0, p3, p5 and p7).}
\end{center}
\end{figure}

The same procedure has been tested on the data obtained in the present simulation for the case of the position B3, where the standard uncertainty on the sample mean value obtained by equation~\eref{eq2} is the largest. The parent population of the statistical variable was obtained generating cosmic ray events corresponding to one year data taking and cumulating them in a single $\Delta x$ distribution, which is shown in~\fref{Fig12}.

This distribution was fitted by an analytical function composed of a sum of four Gaussian functions with the same mean and different standard deviations and weights. 
In~\fref{Fig12} the result of the fitting procedure and the values of the different parameters of the fitting function are shown. 
Then, the same function with only two free parameters, the common mean value and a global scale factor, has been fitted to the twelve samples corresponding to one month data taking. 

In~\tref{Tab1}, for each sample of one month data taking, the following parameters are given: in column~1 the number of the sample; in column~2 the sample mean value; in column~3 the standard uncertainty on the sample mean value given by equation~\eref{eq2}, where $\sigma_{distr}$ is the standard deviation of the sample distribution; in column~4 the mean value of the fitting function used to fit the sample distribution; in column~5 the standard uncertainty on the mean of the fitting function estimated by the fitting procedure. 

In the last two lines of~\tref{Tab1} the following values are reported. 
In the first of them, in column~2 the average value of the mean values on the samples in the same column, in column~3 the average value of the standard errors on the sample mean values in the same column; the error is the standard deviation of the distribution of the values. In columns~4 and~5 the same calculations are performed for the case of the fitting function. 
In the last line, in column~2 the root mean square $s$ of the distribution of the sample mean values in the same column; the uncertainty on the root mean square is calculated by the well known relation $\sigma_{s}$=$s/\sqrt{2(N-1)}$. 
In column~4 the same calculation is performed for the case of the fitting function. 

The results reported in~\tref{Tab1} show that the fluctuation of the mean value of the fitting function used to fit the sample distributions (penultimate line column~5) is at least two time less that the fluctuation of the sample mean (penultimate line column~3). This demonstrate that, with a more refined data analysis, the uncertainty in the measurement of the ``muon target" position could be actually improved by a factor of 2.0 to 3.0.

\begin{table}
\caption{\label{Tab1}For each sample of one month data taking for position B3:  column~2 and 3, mean and standard error on the mean as given by equation~\eref{eq2};  column~4 and 5, mean of the fitting function and standard uncertainty on the mean value estimated by the fitting procedure. In the last line: r.m.s of the values in columns~2 and 4 and average of the values of columns~3 and 5; the errors are calculated as explained in the text.} 
\begin{indented}
\lineup
\item[]\begin{tabular}{@{}*{5}{l}}
\br                              
sample & mean$_{sample}$($\mu$m) & $\sigma_{mean_{sample}}$($\mu$m) & mean$_{fit}$($\mu$m) & $\sigma_{mean_{fit}}$($\mu$m)\cr 
\mr
\0\0 1        &\0\0 \-311  &\0\0 560   &\0\0 \-270   &\0\0  247 \cr 
\0\0 2        &\0\0   368  &\0\0 561   &\0\0   240  &\0\0   245 \cr 
\0\0 3        &\0\0 \-208  &\0\0 565   &\0\0 \-355  &\0\0   246 \cr 
\0\0 4        &\0\0   577  &\0\0 574   &\0\0 \-407  &\0\0   242 \cr 
\0\0 5        &\0\0 \-991  &\0\0 563   &\0\0 \-427  &\0\0   244 \cr 
\0\0 6        &\0\0 \-681  &\0\0 564   &\0\0   26   &\0\0   243 \cr 
\0\0 7        &\0\0 \-117  &\0\0 566   &\0\0 \-99   &\0\0   242 \cr 
\0\0 8        &\0\0   162  &\0\0 564   &\0\0   161  &\0\0   243 \cr 
\0\0 9        &\0\0 \-515  &\0\0 559   &\0\0   148  &\0\0   246 \cr 
\0\0 10       &\0\0 \-279  &\0\0 560   &\0\0 \-371  &\0\0   242 \cr 
\0\0 11       &\0\0 \-1206 &\0\0 566   &\0\0 \-142  &\0\0   250 \cr 
\0\0 12       &\0\0   45   &\0\0 573   &\0\0   418  &\0\0   244 \cr
\mr
\0\0 mean value  &\0\0 \-263         &\0\0  565$\pm$5 & \0\0\ \-90       &\0\0   245$\pm$2  \cr
\0\0 r.m.s.      &\0\0   528$\pm$112 &\0\0            & \0\0  286$\pm$61 &\0\0              \cr
\br
\end{tabular}
\end{indented}
\end{table}

%
\subsection{Measurement of seasonal deformations of the wooden vaulted roof of the ``Palazzo della Loggia"}\label{Muon6}
Owing to the low cosmic ray rate, the proposed monitoring methodology can't provide high precision results in short times. Therefore, it can be competitive with other monitoring techniques only when the deformation under study develops over periods of months or years and the requirement for the monitoring system is to track the slow deformation with time. This is often the case with historical buildings, as in the example illustrated in the present paper. 

In~\fref{Fig4} and~\fref{Fig5}, the cyclic seasonal deformations of few millimeters for two different positions of the wooden vaulted roof structure  of the ``Palazzo della Loggia" are shown. For the position reported in~\fref{Fig5},  the general trend of displacement (green line) shows in addition a rate of deformation of the order of 1~mm per year.

With the aim of demonstrating how a monitoring system based on cosmic ray muon tracking could follow cyclic seasonal deformations as well as systematic ones, a possible seasonal displacement of the three positions B1, B2 and B3 in~\fref{Fig3} was assumed. The behavior of the structure in point B2, reported in \fref{Fig5} for the first year of data taking, was adopted as a realistic model of the phenomenon. Cosmic ray data taking one year long were simulated, with the system monitoring points B1, B2 and B3 of the roof structure.

\begin{figure}
\begin{center}
\includegraphics[width=0.5\textwidth]{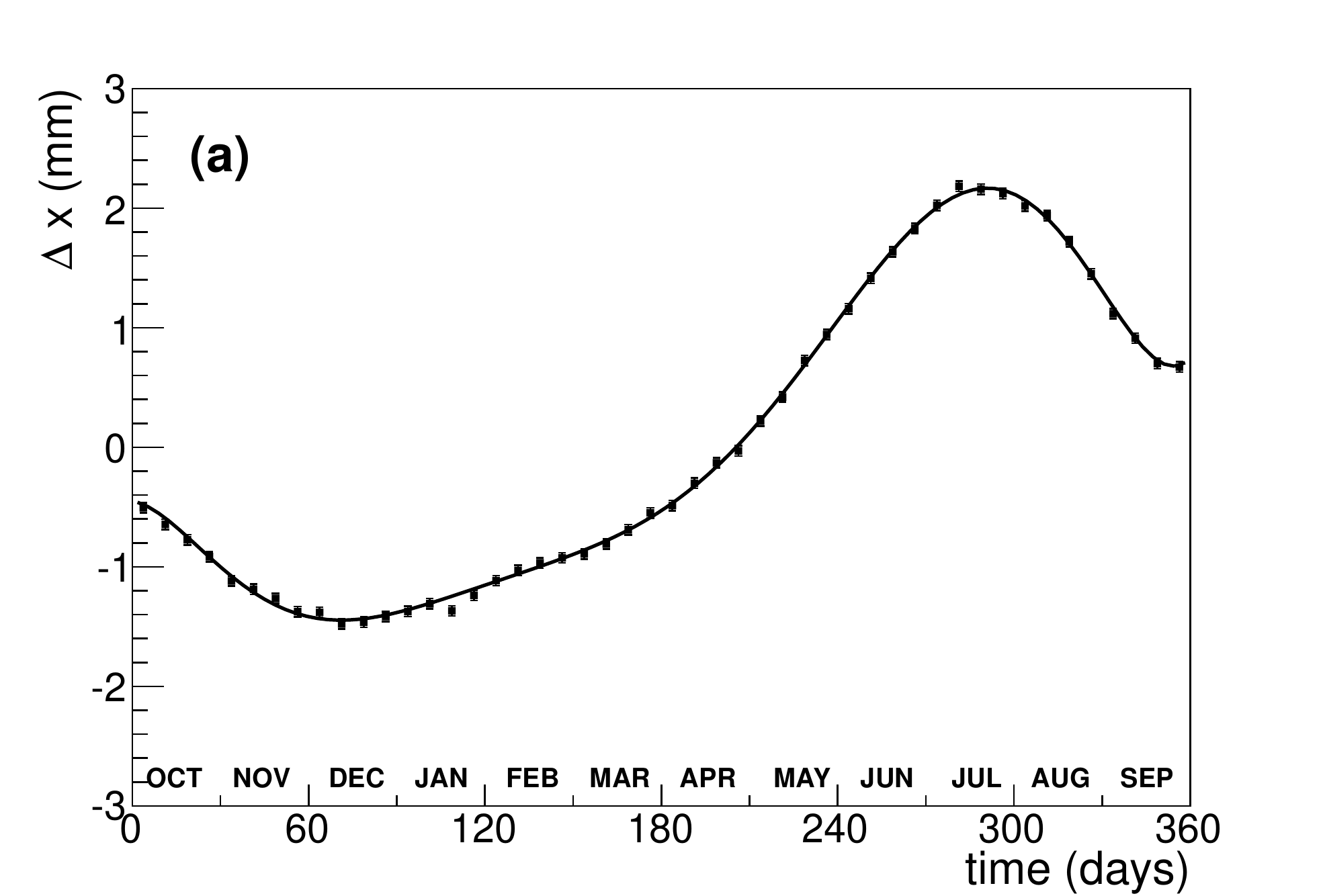}
\includegraphics[width=0.5\textwidth]{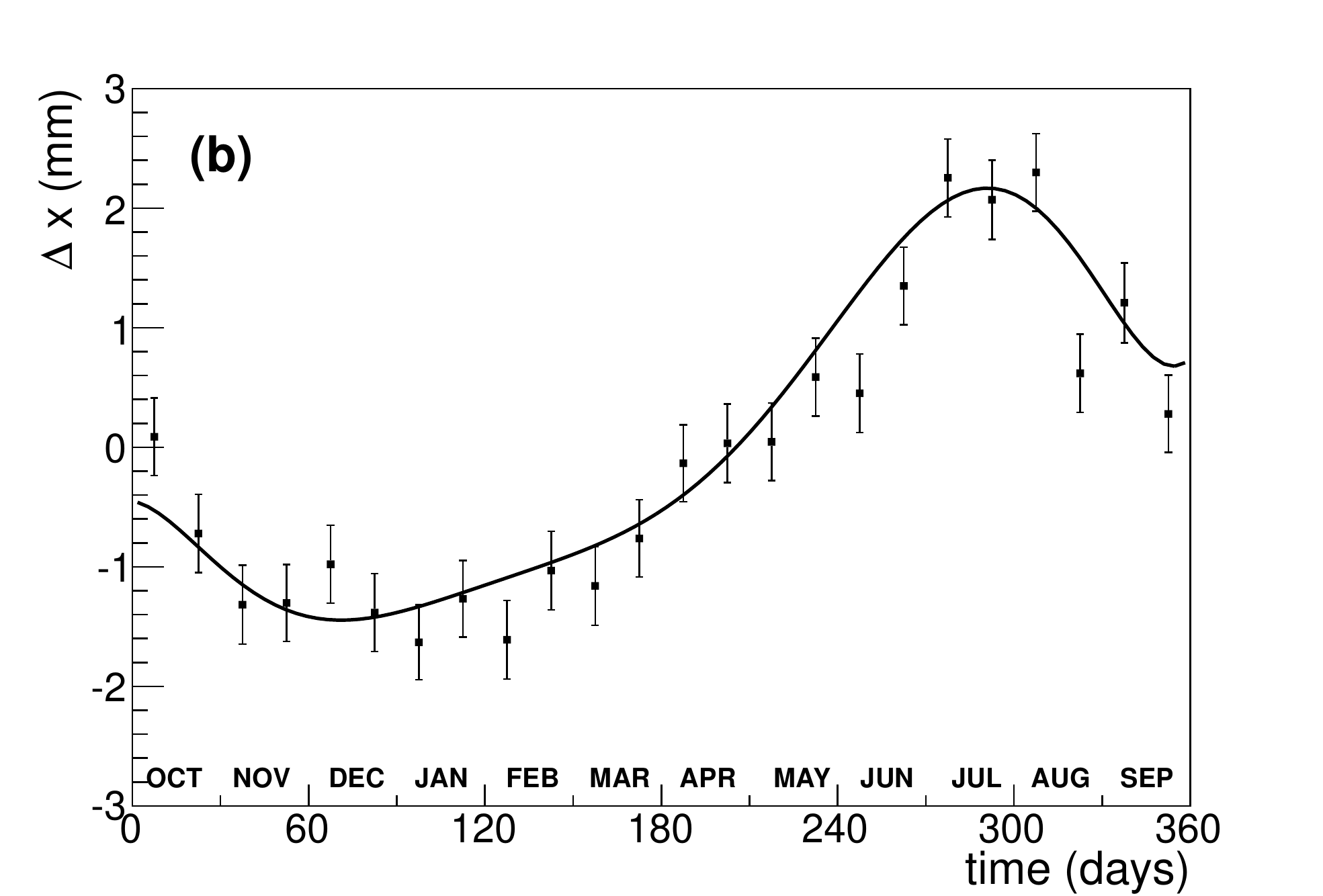}
\includegraphics[width=0.5\textwidth]{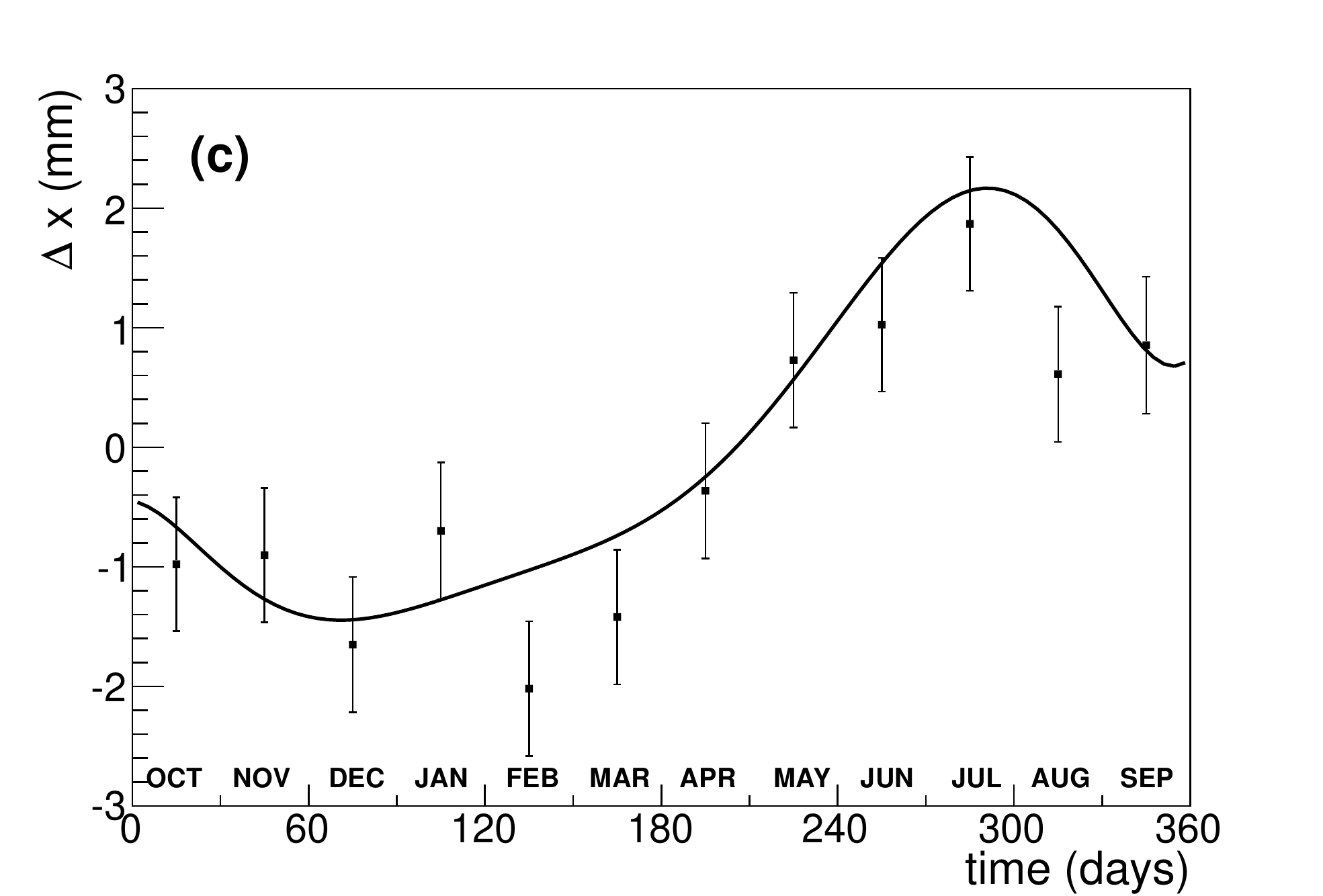}
\caption{\label{Fig13}a) Seasonal deformation of few millimeters assumed for the point in position B1 in~\fref{Fig3} (red line).  Dots indicate the simulated measurement of the position of the ``muon target" with the muon stability monitoring system using the sample mean value, with sampling rate of one week data taking; measurement uncertainty is given by equation~\eref{eq2}. b) The same for the point in position B2, with sampling rate of two weeks. c)  The same for the point in position B3, with sampling rate of one month.}
\end{center}
\end{figure}

For the three points B1, B2 and B3, displacements of the roof structure following the assumed seasonal deformation were simulated. In~\fref{Fig13}, the displacement with time corresponding to the supposed seasonal deformation (the same for the three points) is shown as a continuous line. The results of the simulated measurements of the position of the ``muon target", using the sample mean, displaced following the assumed structural deformation, is shown, with sampling rate of one week, two weeks and one month respectively for points  B1, B2 and B3. 

It is evident the ability of the proposed measurement system to follow seasonal displacements of few millimeters and, even more so, also systematic ones. In addition, as mentioned in sub\sref{Muon5}, with a more refined data analysis, an improvement of the measurement uncertainty of a factor 2 to 3 may be obtained.

%
\section{Comparison with other standard monitoring techniques}\label{Other}
In this paper, the performances of a stability monitoring system, based on the tracking of cosmic ray muons, are compared with the features and performances of the actually adopted monitoring system, based on the measurements of the elongation of metallic wires stretched across the wooden vaulted roof of the ``Palazzo della Loggia".

The latter system has certainly demonstrated, during several years of operation, robustness, good measurement precision, promptness, reliability, at the price of complexity of installation and maintenance (up to 25~m wires to stretch across the roof structure), need of large void spaces to be maintained inaccessible during the measurement, sensitivity to mechanical disturbance and variable climatic conditions, invasiveness, lack of flexibility. 

The former system, in spite of serious limitations in promptness and some limitations in measurement precision, may offer several advantages: (i)~possibility to operate also when the parts to be monitored are not reciprocally visible and are separated by solid masonry structures, as walls and floors; (ii)~flexibility of installation: ``muon telescope" and ``muon target" can be easily located and easily moved and removed, detector disposition may be designed in relation to the specifying monitoring problem; (iii)~absence of moving mechanical parts; (iv)~compactness and absence of mechanical maintenance; (v)~use of a natural, high penetrating radiation, no need of artificial radiation sources and radiation protection safety instructions; (vi)~little invasiveness: no structural modifications requested; (vii)~low sensitivity to changes in environmental conditions, in particular to thermal variations; (viii)~possibility of global simultaneous monitoring of large parts of the building structure.

These particular features may make the proposed system complementary, or even competitive for particular applications, also with other monitoring techniques typically used in this field. Theodolites, laser scanners and LIDAR~\cite{Alba2006,Leica} provide accurate global position measurements, but need that the parts to be monitored be reciprocally visible in a clean atmosphere without turbulence, and request the presence of dedicated operators during the carrying-out of the measurement. Moreover, they are not suitable for data taking over long periods of time.

Pendulums, inclinometers and extensometers~\cite{Doeblin1990} provide deformation or strain measurements in specific point positions; these measurements must be correlated by a model of the building structure, often difficult to obtain accurately in historical buildings. Global position system based methods~\cite{GPS} are hardly applicable to monitor
internal parts of the building and need elaborate post processing of the data to perform high resolution position measurements.

A monitoring system based on cosmic ray muon tracking needs, conversely, the availability of a system of muon detectors with specific features and performances. In next~\Sref{Detector}, the expected characteristics of these detectors are defined and a possible good candidate, out of the many possibilities offered by the nuclear ad particle physics technologies, is identified.

%
\section{Particle detectors suited for a monitoring system based on the tracking of cosmic ray muons}\label{Detector}
The design of a monitoring system based on cosmic ray muon tracking, in the form described in~\Sref{Muon} or in other possible configurations,  needs the availability of muon detector modules featuring characteristics suitable to satisfy the requirements of the specific application. The monitoring system may be particularly suited for applications in which the deformation under study develops over long times and must operate, for long periods, in weakly controlled conditions with low upkeep and steady data acquisition.

Therefore, the muon detectors should necessarily feature the following characteristics: (i)~robustness and stability, insensitiveness to temperature and environmental modifications and limited need of maintenance by qualified personnel; (ii)~structural simplicity, compactness and low weight, with no use of high voltages and gas flushing, in order to simplify the operations of installation of the modules on the field, limit the invasiveness and avoid any mechanical, electrical and chemical hazard; (iii)~high detection efficiency to maximize the acquisition rate and minimize the data taking time needed; (iv)~low power consumption, all on board electronics and totally wireless communication systems to avoid, as much as possible, cabling and separate control units; (v)~use of standard technologies for the particle detection, electronics and data acquisition and, finally, overall low cost. 

Conversely, as shown in~\Sref{Muon} and pointed out in \cite{Bodini2007}, the intrinsic position resolution requested to the muon detectors can be rather coarse. In the present case, a pitch (sensitivity) of few millimeters for the measurement of the coordinates of the muon crossing point on the detector sensitive surface is enough, since the position measurement uncertainty is dominated by stochastic deviations of the muon trajectories.

Possible good detector candidates featuring the requested characteristics are plastic scintillating fibers of square cross section 3.0~mm~$\times$~3.0~mm read by Silicon Photomultipliers SiPM. The use of plastic scintillators, often with wavelength shifting fibers embedded read by segmented cathode photomultipliers, is common in muon radiography and muon tomography applications. In fact, it offers the possibility to cover large surfaces with high detection efficiency using well established technologies at reasonable cost, though with rough position resolution of the order of a few millimeters. 

In~\cite{Tanaka2003,Tanaka2005,Anastasio2013a,Anastasio2013b,Gibert2010}, muon radiography was applied to probe the internal structure of volcanoes. In~\cite{Tanaka2003,Tanaka2005} the inspections of the craters of Mt.~Asama and Mt.~West Iwate in Japan were performed. In the two projects Mu-Ray~\cite{Anastasio2013a,Anastasio2013b} and  DIAPHANE~\cite{Gibert2010}, the technique was applied for the inspection respectively of the Vesuvius, in the gulf of Naples, Italy, and of the Lesser Antilles volcanoes.

In the three previously mentioned cases, the detectors utilized are long scintillator bars with rectangular or triangular cross section, read out by light guides and photomultipliers or, in the two latter cases, by wavelength shifting fibers and photomultipliers or silicon photomultipliers SiPM. The covered detecting surfaces are few square meters. The systems were designed to cope with the extreme working conditions on the field. In~\cite{Tanaka2007a,Marteau2012} also nuclear emulsions are proposed for the detection of the cosmic ray muons for application to the inspection of volcano craters.

In~\cite{Armitage2013}, a Cosmic ray Inspection and Passive Tomography (CRIPT) system has been built. The system covers a 2.0~m $\times$ 2.0~m surface with detector modules formed of two orthogonal layers of scintillator bars, with triangular cross section and 3.3~cm base width, read out by wavelength shifting fibers. The aim of the project is to develop a border security system for the inspection of ULD, unit load devices, standardized shipping containers for aircrafts. 

Within the MuPortal project~\cite{Riggi2013a}, a muon tomography system is under construction, with the aim to inspect the presence of high-Z material hidden in cargo containers. The project foresees the use of muon detector modules formed of plastic scintillator bars 128~cm long and of 1.0~cm~$\times$~1.0~cm cross section, read out by wavelength shifting fibers, to cover a surface of 2.54~m $\times$ 12.80~m.

In muon tomography applications, when very large volumes have to be covered and high spatial resolution precision is requested, the use of drift chambers has been preferred. This is the case of the MuSteel project~\cite{Musteel2010}, aimed at the construction of a muon scanner to detect radioactive orphan sources hidden in scrap metal containers for the steel industry. Indeed, gas detectors are able to cover very large sensitive surfaces and perform position measurement uncertainties of the order of 100~$\mu$m. The total horizontal surface covered by the designed detection system is indeed 18.0~m~$\times$~6.0~m, and the total sensitive surface about 400~m$^2$, in order to allow an entire truck to be completely inspected in few minutes of data taking.

When the volumes to be inspected are smaller, as for industrial nuclear waste containers, the use of detection systems based on layers of scintillating fibers is often proposed. For this particular industrial application, a prototype muon detectors system has been developed and realized~\cite{Mahon2013,Clarkson2013} at the University of Glasgow. The single detector module is composed of two orthogonal layers of 128~plastic scintillanting fibers, 2~mm pitch, for a total surface of 25.6~cm~$\times$~25.6~cm, read by segmented anode photomultipliers.

In this context, the adoption, for the proposed monitoring system, of cosmic ray muon detectors based on plastic scintillator fibers and silicon photomultipliers detectors SiPM seems fully appropriate. A prototype of the ``muon telescope" system, in reduced scale, is at present under construction.

The muon detector module is formed of two orthogonal layers of scintillating fibers 3.0~mm~$\times$~3.0~mm square cross section, 20~cm long, BCF-10 Bicron Saint Gobain, read at one polished side by a silicon photomultiplier SiPM NUV with 3.0~mm~$\times$~3.0~mm sensitive surface by AdvanSiD. The SiPM are made up of multiple APD avalanche photodiode pixels with 50~$\mu$m~$\times$~50~$\mu$m surface operated in Geiger mode. They are essentially opto-semiconductor devices with excellent photon counting capability and with the great advantages of low voltage operation (about 30~V) and insensitivity to magnetic field.

\begin{figure}
\begin{center}
\includegraphics[width=0.7\textwidth]{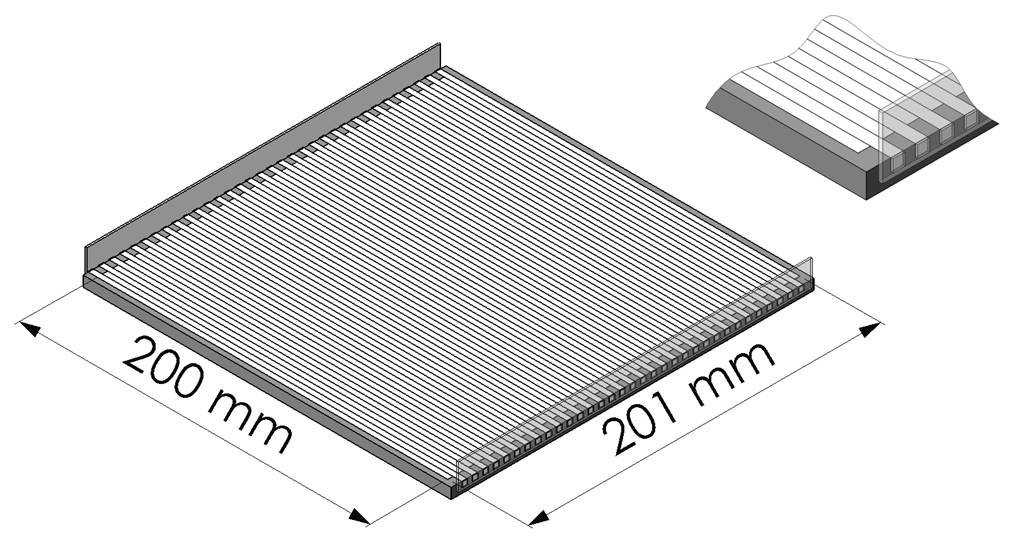}
\caption{\label{Fig15}Assembling scheme of a muon detector layer composed of 67 scintillating fibers 3.0~mm~$\times$~3.0~mm square cross section 20~cm long. Scintillating fibes are read alternatively at the two sides by a SiPM with 3.0~mm~$\times$~3.0~mm sensitive surface. The plastic mechanical support structure and printed circuit boards are shown. }
\end{center}
\end{figure}

The proposed assembling scheme of the single detector layer composing the detector module is shown in~\fref{Fig15}. Sixty scintillating fibers, 3.0~mm~$\times$~3.0~mm square cross section 20~cm long, are assembled in contact one with the other in order to minimize detection inefficiencies. They are read alternatively at the two sides by a SiPM, kept in contact with one polished scintillating fiber side by optical grease.
The cross talk between adjacent fibers can be limited either by the use of appropriate reflector paints for plastic scintillators or by the interposition of a very thin layer of black wrapping tape. For mechanical support and light hermeticity a robust black plastic support structure and cover have been designed.

\begin{figure}
\begin{center}
\includegraphics[width=0.7\textwidth]{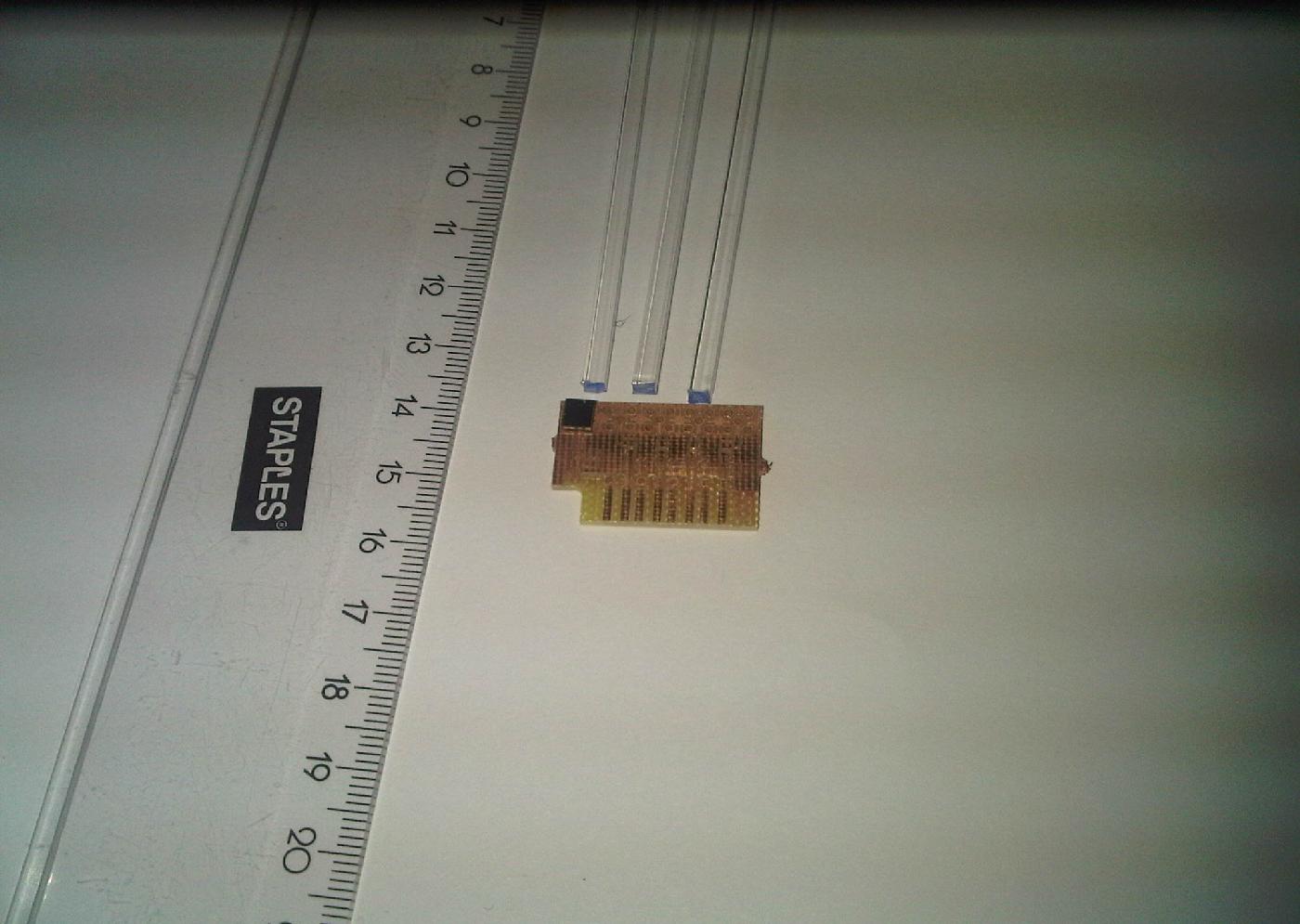}
\caption{\label{Fig16}Square scintillating fibers  3.0~mm~$\times$~3.0~mm square cross section and SiPM detectors mounted on the supporting printed circuit board.}
\end{center}
\end{figure}

\begin{figure}
\begin{center}
\includegraphics[width=0.7\textwidth]{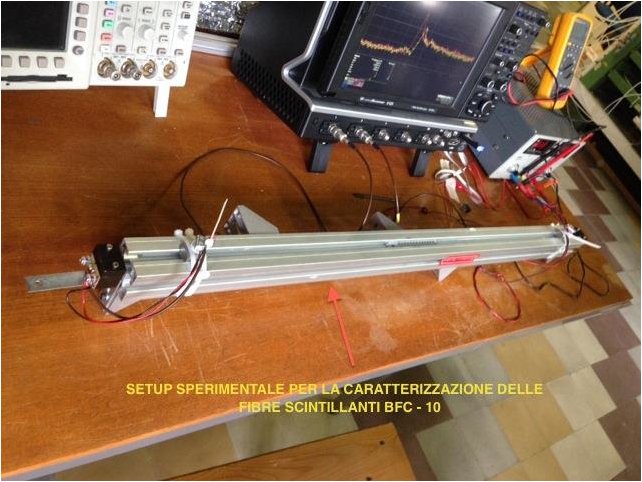}
\caption{\label{Fig17} Testing system of a single square scintillating fiber: the  fiber is one meter long, it is sustained by an aluminum structure and is read on both polished sides by two SiPM photomultipliers.}
\end{center}
\end{figure}

The first experimental tests on the elementary detection modules have been performed. In~\fref{Fig16} a few  square scintillating fibers and SiPM detectors mounted on the supporting printed circuit board are shown. In~\fref{Fig17} the testing system of a single square scintillating fiber is shown: the scintillating fiber is one meter long, it is sustained by an aluminum structure and is read out on both polished sides by two SiPM photomultipliers.

Tests of the amplitude of the signals collected in coincidence at both extremities of a scintillating fiber with a non collimated source of $^{90}$Sr, pure $\beta$ emitter, placed in different positions along the fiber, have been performed. Signal amplitudes of 6~mV average value have been measured on both SiPM detectors without any amplification with timing resolution better then 500~ps. 
These results demonstrate that the sensitive element of the detecting system is able to perform a good detection efficiency and that the electronic chains needed to amplify and the signal for the data acquisition and processing can be rather simple and cheap. Further extensive experimental tests and simulations of the designed detector to define sensitivity, efficiency and cross talk are in progress. The results will be presented in a forthcoming paper devoted to the design of the full detector system and assessment of its performances.

%
\section{Summary and conclusions}\label{Summary}
Cosmic ray radiation, known since the first decades of the 20$^{th}$ century, has been widely used in the field of nuclear and particle physics as a source of high energy projectiles for the investigations of the fundamental laws of nature and as a tool, naturally available at the Earth surface, for particle detector testing and calibration and for detector mechanical alignment in complex apparatus. 

Thanks to the high penetrability of cosmic ray radiation, it has been also applied in fields different from nuclear and particle physics as geological research, archaeological studies, industrial and civil security applications for the inspection and imaging of the content of large, dense and inaccessible volumes. The principal techniques utilized are muon radiography and muon tomography, the latter particularly in the search for high-Z materials in cargoes and containers.

Recently, it has been suggested that, due to its capability of crossing very thick layers of material suffering only small deviations, cosmic ray radiation may be used for measurement applications in mechanical and civil engineering, with specific reference to situations where environmental conditions are weakly controlled and/or when the parts to be monitored are not mutually visible. 
In the present paper, the application of the muon stability monitoring method to historical buildings has been studied, by Monte Carlo technique. A realistic situation was considered: the exemplary case of the wooden vaulted roof of the ``Palazzo della Loggia" in the town of Brescia, for which a stability monitoring campaign was performed, for more than ten years, by means of traditional mechanical methods.

A measurement system formed of a ``muon telescope", to be located on a fixed part of the building (the reference system) and by a ``muon target", to be positioned in the part of the building whose position has to be monitored, has been designed.
The ``muon telescope" is composed of three muon detector modules, axially aligned 50~cm apart from each other, whose sensitive volume is a square 40~cm $\times$ 36~cm side and 6.0~mm thickness, made of two orthogonal layers of square scintillating fibers 3.0~mm~$\times$~3.0~mm cross section. The ``muon target" is composed of one detector module of the same size.

In a Monte Carlo calculation program based on GEANT4 toolkit, the proposed muon stability monitoring system, the geometry and relevant structure of the building and a realistic cosmic ray muon generator have been modeled. The procedure of measurement of the positions of three different points in the wooden vaulted roof, relative to the fixed reference system separated by a bulky wooden ceiling 15~cm thick, has been simulated.

Position measurement uncertainties of the designed muon monitoring system have been calculated as a function of the data taking time. The calculations demonstrated that the designed system may perform measurement precisions consistent with the amount of displacements under observation, with data taking times compatible with the time scale characteristic of the deformation phenomenon.

Both cyclic and systematic displacements observed in the ``Palazzo della Loggia" during several years of observation could have been observed, with suitable precision of less than 1.0~mm, by a monitoring system based on the tracking of cosmic ray muons.
In addition, it was pointed out that the efficiency of the designed system, and consequently data taking times, may easily be improved of a factor 2 to 3  by improving the data analysis. Consistent improvements in performances can also be obtained  by modifying some geometrical parameters of the proposed measurement system.

In conclusion, cosmic ray muon detection techniques are assessed for measurement applications in the field of civil engineering and may be particularly suitable for static monitoring of historical buildings, where the evolution of the deformation phenomena under study is of the order of months or years. Appealing features of the proposed monitoring system are: (i)~the use of a natural and ubiquitous source of radiation; (ii)~the applicability also in presence of horizontal and/or vertical building structures interposed between the reference system and the parts to be monitored; (iii)~the limited invasiveness, and the flexibility and easiness of installation of the monitoring system devices; (iv)~the possibility to design a global monitoring system, where the position of different points of the building may be  simultaneously monitored relative to the same reference system; (v)~the use of well known physical principles and established technologies in the field of nuclear and particle physics. 

Limiting features are the intrinsic stochastic nature of the behavior of the radiation utilized, which requests to cumulate statistical distributions to be treated by statistical inference methods, and the low rate of cosmic ray radiation, which makes this technique generally unfit for applications where promptness of response is requested. The performances of such measurement system strongly depend on the particular application under study, geometries and interposed materials. However, the system performances in the specific situation may be easily evaluated by Monte Carlo calculations and the system can be designed accordingly.

The availability of muon detector modules featuring characteristics suitable to satisfy the specific application requirements is essential for the described technique to be realistically proposed. Scintillating fibers of square cross section few millimeter side read by silicon photomultipliers SiPM appear as very promising candidates, coping well the requirements of robustness, efficiency, stability and reliability, absence of any hazard, low cost that such a system should necessarily perform to be proposed for potential applications.

%
\ack
The authors gratefully acknowledge Prof. Ezio Giuriani and Prof. Alessandra Marini of the Department of Civil, Architectural, Land and Environmental Engineering and Mathematics of the Brescia University for the information provided on the long-lasting study performed of the ``Palazzo della Loggia" and for the invaluable advice concerning the problem of historical building monitoring. 
This work has been done thanks to a special funding of the Department of Mechanical and Industrial Engineering of the University of Brescia.

%

\end{document}